\documentclass[useAMS,usenatbib]{mn2e} 
\usepackage{url,graphics,times,array,amsmath,amssymb,ctable,enumitem}

\newcommand{\CPDF}{P(<\tau_{\mathrm{eff}})}
\newcommand{\Mpc}{\mathrm{Mpc}}
\newcommand{\kpc}{\mathrm{kpc}}
\newcommand{\MFP}{\lambda_{\mathrm{mfp}}^{912}}
\newcommand{\Mmin}{M_{\mathrm{min}}}
\newcommand{\Msun}{M_{\odot}}
\newcommand{\GammaHI}{\Gamma_{\mathrm{HI}}}

\newcommand{\dd}{\mathrm{d}}
\newcommand{\taueff}{\tau_{\mathrm{eff}}}

\newcommand{\HI}{H{\sc~i}}

\newcommand{\HeI}{He{\sc~i}}
\newcommand{\HeII}{He{\sc~ii}}

\newcommand{\addition}{}

\title[Ionizing Background Fluctuations]{Large Fluctuations in the High-Redshift Metagalactic Ionizing Background}

\author[D'Aloisio et al.]{Anson D'Aloisio$^1$\thanks{Email:anson@u.washington.edu}, 
Matthew McQuinn$^1$, Frederick B. Davies$^2$, and Steven R. Furlanetto$^3$ \smallskip \\ 
$^1$Astronomy Department, University of Washington, Seattle, WA 98195, USA \\
$^2$Max Planck Institut f\"{u}r Astronomie, K\"{o}nigstuhl 17, D-69117, Heidelberg, Germany\\
$^3$Department of Physics and Astronomy, University of California, Los Angeles, Box 951547, Los Angeles, CA 90095, USA }

\begin{document}
\maketitle

\begin{abstract}
Recent observations have shown that the scatter in opacities among coeval segments of the Ly$\alpha$ forest increases rapidly at $z > 5$.  In this paper, we assess whether the large scatter can be explained by fluctuations in the ionizing background in the post-reionization intergalactic medium.  We find that matching the observed scatter at $z\approx 5.5$ requires a short spatially averaged mean free path of $\langle \MFP \rangle \lesssim 15 h^{-1}$ comoving Mpc, a factor of $\gtrsim 3$ shorter than direct measurements at $z = 5.2$.  We argue that such rapid evolution in the mean free path is difficult to reconcile with our measurements of the global \HI\ photoionization rate, which stay  approximately constant over the interval $z \approx 4.8 - 5.5$.    However, we also show that measurements of the mean free path at $z>5$ are likely biased towards higher values by the quasar proximity effect.  This bias can reconcile the short values of $\langle \MFP \rangle$ that are required to explain the large scatter in opacities.   We discuss the implications of this scenario for cosmological reionization.  Finally, we investigate whether other statistics applied to the $z>5$ Ly$\alpha$ forest can shed light on the origin of the scatter. Compared to a model with a uniform ionizing background, models that successfully account for the scatter lead to enhanced power in the line-of-sight flux power spectrum on scales $k \lesssim 0.1~h \Mpc^{-1}$.  We find tentative evidence for this enhancement in observations of the high-redshift Ly$\alpha$ forest.   
\end{abstract}


\section{Introduction}

The high-redshift Ly$\alpha$ forest provides our most robust observational constraints on the end of the cosmological reionization process.  The fact that most $z < 6$ segments of the forest show any transmission indicates that reionization was largely complete by $z \approx 6$, when the Universe was one billion years old  \citep{fan06, 2008MNRAS.386..359G, mcgreer15}.  In addition, Ly$\alpha$ forest measurements of the metagalactic ionizing background strength \citep[e.g.][]{2002AJ....123.1247F, 2007MNRAS.382..325B, 2013MNRAS.436.1023B} and of the thermal state of the intergalactic medium (IGM) \citep[e.g.][]{2000MNRAS.318..817S, 2010ApJ...718..199L, 2011MNRAS.410.1096B,2014MNRAS.441.1916B} serve as important boundary conditions for theoretical models of reionization (see the review of \citealt{2015arXiv151200086M}, and references therein).   

The opacity of the Ly$\alpha$ forest is often quantified by the effective optical depth, $\taueff=-\ln\langle F\rangle_L$.  Here, $F \equiv \exp(- \tau_{\mathrm{Ly} \alpha} )$ is the transmitted fraction of the quasar's flux, $\tau_{\mathrm{Ly}\alpha}$ is the optical depth in Ly$\alpha$, and $\langle \ldots \rangle_L$ denotes a line-of-sight average over a segment of the forest of length $L$.  The Ly$\alpha$ optical depth scales as $\tau_{\mathrm{Ly}\alpha} \propto T^{-0.7} \Delta_b^2/\GammaHI$, where $T$ is the gas temperature, $\Delta_b$ is the gas density in units of the cosmic mean, and $\GammaHI$ is the \HI\ photoionization rate (which scales with the strength of the local ionizing background).  Previous studies have noted a steep increase in both the mean of $\taueff$ and its dispersion among coeval segments of the forest around $z=6$.  Some authors have interpreted these trends as signatures of the last stages of reionization \citep{2000ApJ...535..530G, fan06, 2015MNRAS.447.3402B}.   Whether the quick evolution in the mean of $\taueff$ signifies the end of reionization is a topic of debate, however, as rapid evolution may be possible even after reionization \citep{lidz06,2011ApJ...743...82M, munoz14}.  

Evolution in the dispersion of $\taueff$ may hold more promise for studying reionization.  Improving upon previous measurements, \citet{2015MNRAS.447.3402B} showed that the dispersion in $\taueff$ amongst coeval $L=50h^{-1}~\Mpc$ segments of the $z\gtrsim 5.5$ forest far exceeds the expected dispersion from density fluctuations alone.   They attributed this finding to the presence of large spatial fluctuations in the ionizing background, indicative of the final stages of reionization.  \citet{2015ApJ...813L..38D} proposed an alternative scenario in which the excess $\taueff$ dispersion is generated by residual temperature inhomogeneities in the IGM, imprinted by the patchy reionization process.  In their model, the observed amplitude of $\taueff$ fluctuations implies a late-ending ($z\approx 6$) and extended reionization process, with roughly half of the volume reionized at $z \gtrsim 9$.  While such large-scale temperature variations, if confirmed, would be a unique signature of patchy reionization, it is important to note that strong fluctuations in the ionizing background do not necessarily indicate the final stages of reionization.  They could be of a simpler origin, reflecting instead the abundance and clustering of the sources, as well as the mean free path of \HI\ ionizing photons, $\MFP$, through the \emph{post-reionization} IGM.  The latter sets the spatial scale above which $\GammaHI$ is expected to fluctuate significantly.     

Models of the post-reionization ionizing background typically assume that stellar emissions from galaxies are the dominant source of ionizing photons at $z\gtrsim 4$ \citep[e.g.][]{2009ApJ...703.1416F,2012ApJ...746..125H}.  This assumption is based on measurements of the mean transmission in the Ly$\alpha$ forest and the rapid decrease in the quasar abundance observed at $z>3$ \citep[e.g.][]{2007MNRAS.382..325B, 2013MNRAS.436.1023B, 2006AJ....132..117F,2010AJ....139..906W,2013ApJ...768..105M,2015MNRAS.453.1946G}. Most models have adopted the approximation of a spatially uniform $\MFP$, with values guided by direct measurements of the mean free path from quasar absorption spectra \citep{2009ApJ...705L.113P,2014MNRAS.445.1745W}.  However, this approximation is unlikely to be valid on the scales of interest for the $z>5$ Ly$\alpha$ forest.  Using cosmological hydrodynamics simulations post-processed with radiative transfer, \citet{2011ApJ...743...82M} showed that the mean free path should vary with the strength of the local background as $\MFP \propto \GammaHI^{2/3 - 3/4}$.  These variations reflect the enhancement (suppression) of self-shielded absorbers in regions where the background is weaker (stronger).  This mutual feedback between $\MFP$ and $\GammaHI$ amplifies fluctuations in $\GammaHI$ over a model with uniform $\MFP$. 

At present, it is prohibitively expensive to capture these effects in fully numerical simulations, as doing so requires resolving the absorbers that regulate $\MFP$ {\it and} sampling the cosmological scales ($L \gg \MFP$) over which $\GammaHI$ varies. To bridge this gap, \citet{2015arXiv150907131D} developed a semi-numerical model of the post-reionization ionizing background that accounts for the effects of variations in $\MFP$ over cosmologically representative scales.  The amplitude of $\taueff$ fluctuations in their model depends on the spatial average of the mean free path, $\langle \MFP \rangle$, with smaller $\langle \MFP \rangle$ leading to larger fluctuations. They found that the $z\approx 5.5$ opacity fluctuations observed by \citet{2015MNRAS.447.3402B} favor a short value of $\langle \MFP \rangle \approx 10 h^{-1}$ (comoving) $\Mpc$.  This scenario leads to a very different opacity structure than the low-redshift forest; regions that are underdense in sources, i.e. cosmic voids, are the most \emph{opaque} (highest $\taueff$) $L=50~h^{-1}\Mpc$ segments of the forest, while density peaks rich with sources are the most transmissive (lower $\taueff$).  


The small values of $\langle \MFP \rangle$ that seem necessary to account for the observations of \citet{2015MNRAS.447.3402B} have led some authors to question whether galaxies are actually the dominant sources of the $z>5$ ionizing background.  \citet{2015arXiv150501853C} proposed a model in which the dispersion is driven primarily by the rarity and brightness of the sources. In this model, active galactic nuclei (AGN) account for a significant fraction ($\gtrsim 50\%$) of the \HI\ ionizing emissions at $z>5$ \citep{2016arXiv160608231C}.  We explore these AGN-driven models further in a companion paper (\citealt{2016arXiv160706467D}, henceforth ``Paper II").  There we argue that the photoheating from the earlier onset of \HeII\ reionization in these models is in tension with recent measurements of the IGM temperature.   

In this paper we will assess the model of \citet{2015arXiv150907131D} under the standard assumption that galaxies are the dominant sources.  Since the background fluctuations in this model are generated entirely by the interplay between galaxy clustering and large-scale variations in the mean free path, we will effectively assume that reionization concludes early enough for gas relaxation effects and residual temperature fluctuations from reionization to be negligible.\footnote{We will, however, discuss the implications of these assumptions below.}  Whereas the calculations of \citet{2015arXiv150907131D} were based on semi-numerical simulations of structure formation and an approximate model for the opacity of the Ly$\alpha$ forest, we will utilize mock absorption spectra extracted from a cosmological hydrodynamics simulation with $2\times2048^3$ resolution elements.  

Foreshadowing, in \S \ref{SEC:intensityflucs} we confirm that matching the large $z\approx 5.5$ opacity variations in \citet{2015MNRAS.447.3402B} requires that $\langle \MFP \rangle \lesssim 15 h^{-1} \Mpc$.  This is a factor of $\gtrsim 3$ shorter than the direct measurement of \citet{2014MNRAS.445.1745W}, $\MFP = 44 \pm 7~h^{-1} \Mpc$ at $z\approx 5.2$ (the highest redshift currently available).  In \S \ref{SEC:globalmeasurements}, we argue that such a rapid evolution in the mean free path is difficult to reconcile with the remarkably flat evolution in the global photoionization rate, as measured from the mean flux of the Ly$\alpha$ forest. However, in \S \ref{SEC:MFPbias}, we show that the quasar proximity effect can bias measurements of the mean free path higher than $\langle \MFP \rangle$ by up to a factor of $\approx 2$.  This bias can reconcile the measurements with the short values of $\langle \MFP \rangle$ that are required to explain the large opacity variations.  We discuss the implications of this scenario for observations of the $z>5$ Ly$\alpha$ forest and for cosmological reionization. In \S \ref{SEC:fluxpower}, we explore the use of other statistics applied to the Ly$\alpha$ forest to probe the large opacity fluctuations. Finally, in \S \ref{SEC:conclusion} we offer concluding remarks.  From here on all distances are reported in comoving units unless otherwise noted.  For all computations we assume a vanilla $\Lambda$CDM cosmology with $\Omega_m=0.31$, $\Omega_b=0.048$, $h=0.68$, $\sigma_8=0.82$, $n_s=0.97$, and $Y_{\mathrm{He}}=0.25$, consistent with recent measurements \citep{2015arXiv150201589P}.

 \section{Opacity Fluctuations in the High-Redshift Ly$\alpha$ Forest}
 \label{SEC:intensityflucs}  
 
 \subsection{Numerical methodology}
 \label{SEC:methodology}

\begin{figure}
\begin{center}
\resizebox{8.4cm}{!}{\includegraphics{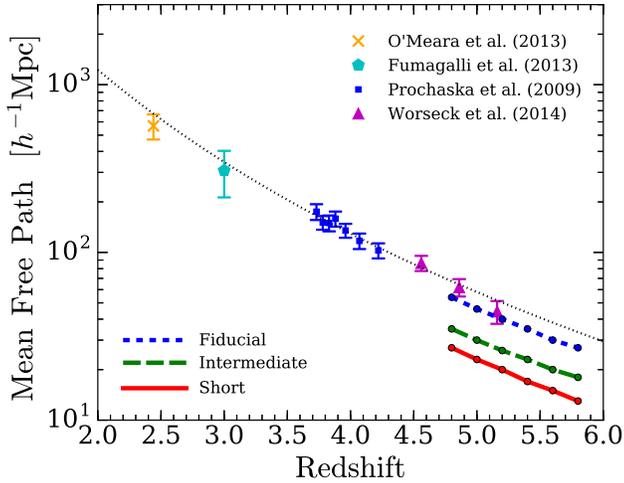}}
\end{center}
\caption{The three models for $\langle \MFP \rangle$ used in this paper (thick curves).   The blue/short-dashed, green/long-dashed, and red/solid curves correspond to our ``fiducial," ``intermediate," and ``short" mean free path models, respectively.  Values of $\langle \MFP \rangle$ in these models are reported in Table \ref{TAB:MFPmodels}. The data points are measurements, while the thin black/dotted curve shows the empirical fit of \citet{2014MNRAS.445.1745W}.  }  
\label{FIG:MFPmodels}
\end{figure}

To model the galaxy population, as well as the density and temperature structure of the Ly$\alpha$ forest, we ran a cosmological hydrodynamics simulation using a modified version of the { Eulerian} hydro code of \citet{2004NewA....9..443T}, with $N_{\mathrm{dm}}=2048^3$ dark matter particles and $N_{\mathrm{gas}}=2048^3$ gas cells, in a periodic box with a side length of $L_{\mathrm{box}}=200h^{-1}~\Mpc$.  These parameters strike a balance between the large volumes required to capture large-scale fluctuations in the ionizing background \citep{2015arXiv150907131D}, and the high resolutions required to model the high-$z$ Ly$\alpha$ forest \citep{2009MNRAS.398L..26B}.  { We use a primordial power spectrum generated with the CAMB software \citep{Lewis:1999bs}, and we initialize the simulation at $z=300$ using 1st order Lagrangian Perturbation Theory.  Our simulation does not include the effects of galaxy formation -- e.g. star formation, supernovae and AGN feedback, and metal enrichment -- but these mechanisms are not expected to have a significant impact on the low-density gas probed by the $z>5$ Ly$\alpha$ forest \citep[e.g.][]{2015arXiv150501853C}.}

{ The reionization of \HI\ is modeled in a simplistic way by instantaneously turning on a uniform ionizing background (with $\GammaHI = 10^{-13}$ s$^{-1}$) and heating the box uniformly to a temperature of $T=20,000~\mathrm{K}$ at $z=7.5$ (see \citealt{2015ApJ...813L..38D} for more details on our simulation methodology).  This redshift of reionization was chosen such that the thermal history of the simulation approximately matches the $z\approx 4.8$ temperature measurement of \citet{2011MNRAS.410.1096B} (see Fig. \ref{FIG:sim_thermal_hist}).  Following the instantaneous reionization heating, our simulation tracks the thermal history of the gas in the presence of a uniform ionizing background, with a nominal normalization of $\GammaHI = 10^{-13}$ s${^{-1}}$, and a power law spectrum, $J_{\nu} \propto \nu^{-\alpha_{\mathrm{bkgd}}}$, with $\alpha_{\mathrm{bkgd}} = 0.5$.  Below, we describe how we rescale the background normalization under the assumption of photoionization equilibrium to match the observed opacity of the Ly$\alpha$ forest.  We include photoheating from \HI\ and \HeI\ ionizations and all of the relevant cooling processes for a gas of primordial composition (Compton, recombination, free-free, collisional ionization, and collisional excitation). }  Using smaller test simulations, we find that the distribution of $\taueff$ fluctuations is insensitive to the particular \HI\ reionization temperature and redshift in the absence of relic temperature fluctuations from patchy reionization, which the simulations here do not attempt to model.   We assume that photoheating from the reionization of \HeII\ is negligible at the redshifts of interest for this paper ($z>5$), consistent with standard models of \HeII\ reionization \citep[see][and references therein]{2015arXiv151200086M}. 
 
Halos were identified on-the-fly using a spherical over-density criterion in which the enclosed mass is $M_{200} = (4 \pi/3) 200\bar{\rho}_m R_{200}^3$, where $\bar{\rho}_m$ is the cosmic mean matter density, and $R_{200}$ is the radius below which the mean over-density of the halo is $> 200 \bar{\rho}_m$.  For galaxies we abundance match the halo catalogs to an extrapolation of the luminosity function measured by \citet{2015ApJ...803...34B} (see \citet{2015ApJ...813...54T} for more details on the halo finder and the abundance matching scheme).  We assume that all halos with masses above $\Mmin=2\times10^{10}h^{-1}~\Msun$ host a star-forming galaxy, which corresponds to a lower magnitude limit of $M_{\mathrm{AB},1600}\approx-17.5$ at $z=5.5$. This minimum threshold of $\approx300$ dark matter particles per halo was chosen for completeness of the halo mass function.  We note that the $z\approx 5$ luminosity function measurement of \citet{2015ApJ...803...34B} extends down to $M_{\mathrm{AB},1600}\approx-16.4$, such that the sources in our simulations are above detection limits. {\addition For reference, the number of sources in our simulation volume at $z=5.6$ is $210,823$, and the brightest source has a UV magnitude of $M_{\mathrm{AB},1600}=-23.2$.  Including less clustered halos down to a lower mass limit (e.g. the $\Mmin \sim 10^9h^{-1}~\Msun$ threshold assumed in \citealt{2015arXiv150907131D}) would have the impact of reducing the amplitude of ionizing background fluctuations in our models.  We return to this point below. }


Following \citet{2015MNRAS.447.3402B} and \citet{2015arXiv150907131D}, we assume a flat spectrum at $\lambda>912$ \AA, with a break of $f_{\rm esc}/A_{912}$ at $912$ \AA, where $A_{912} = L_{\nu}(1600\mathrm{\AA})/L_{\nu}(912 \mathrm{\AA})$ represents the Lyman break from absorption in stellar atmospheres, and $f_{\rm esc}$ is a free parameter representing the escape fraction of \HI\ ionizing radiation into the IGM.  ($f_{\rm esc}$ also effectively encapsulates uncertainties in $A_{912}$.)  We adopt a nominal value of $A_{912}=6$, representative of the expected value for young stellar populations, but we note that $A_{912}$ can be as low as $2-3$ (\citealt{1999ApJS..123....3L}; \citealt{2003MNRAS.344.1000B}; \citealt{2012MNRAS.419..479E}; see Fig. 1 of \citealt{2007ApJ...668...62S}).  We further assume that $L_{\nu}\propto\nu^{-2}$ for $\lambda\leq912$ \AA.   None of these assumptions affect our main conclusions because the \HI\ photoionization cross section scales steeply with frequency ($\sigma_{\rm{HI}} \propto \nu^{-2.8} $), such that most of $\GammaHI$ owes to photons with wavelengths just below $\lambda = 912$ \AA.  

\begin{figure*}
\begin{center}
\resizebox{6.0cm}{!}{\includegraphics{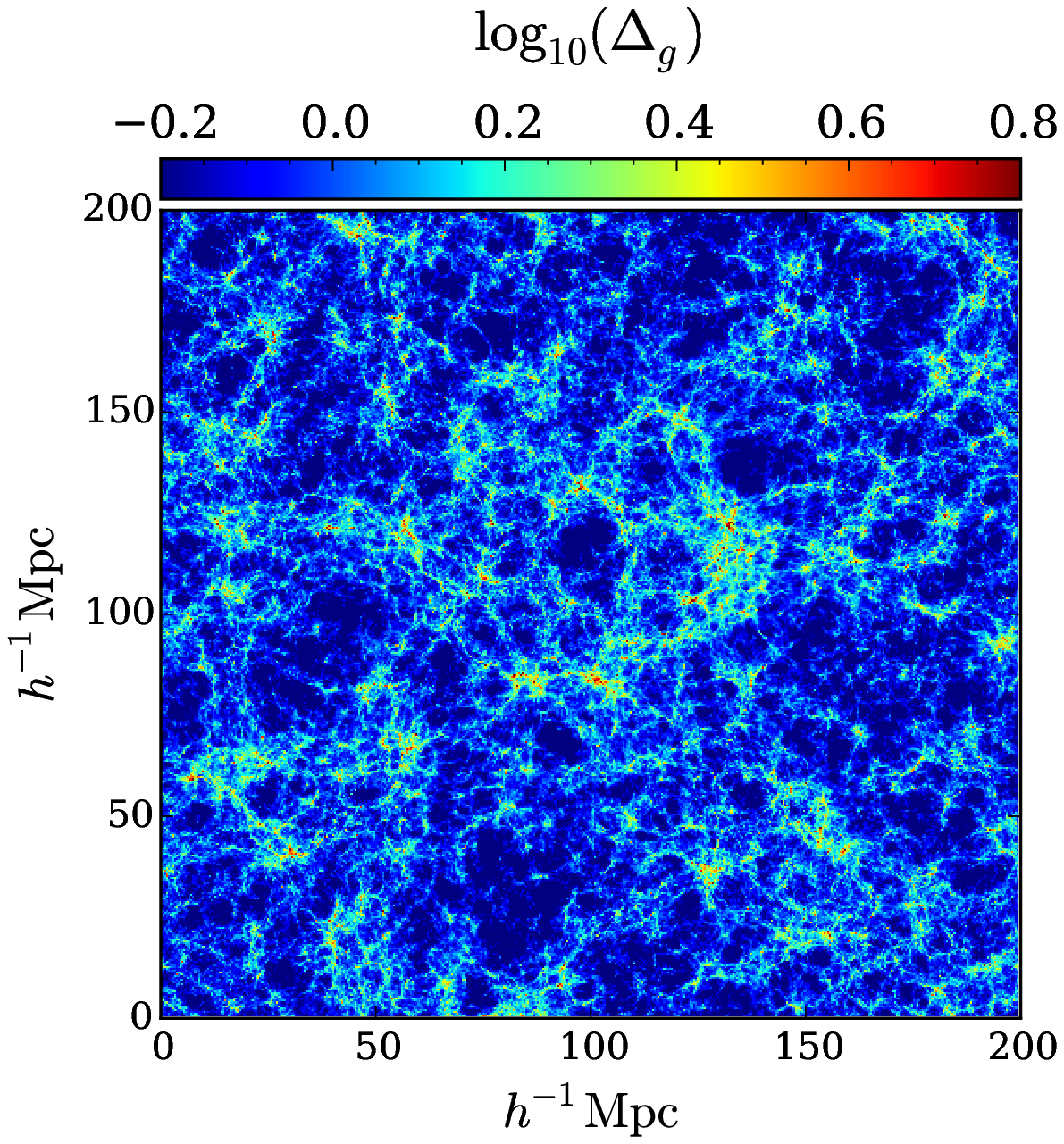}}
\hspace{-0.28cm}
\resizebox{6.0cm}{!}{\includegraphics{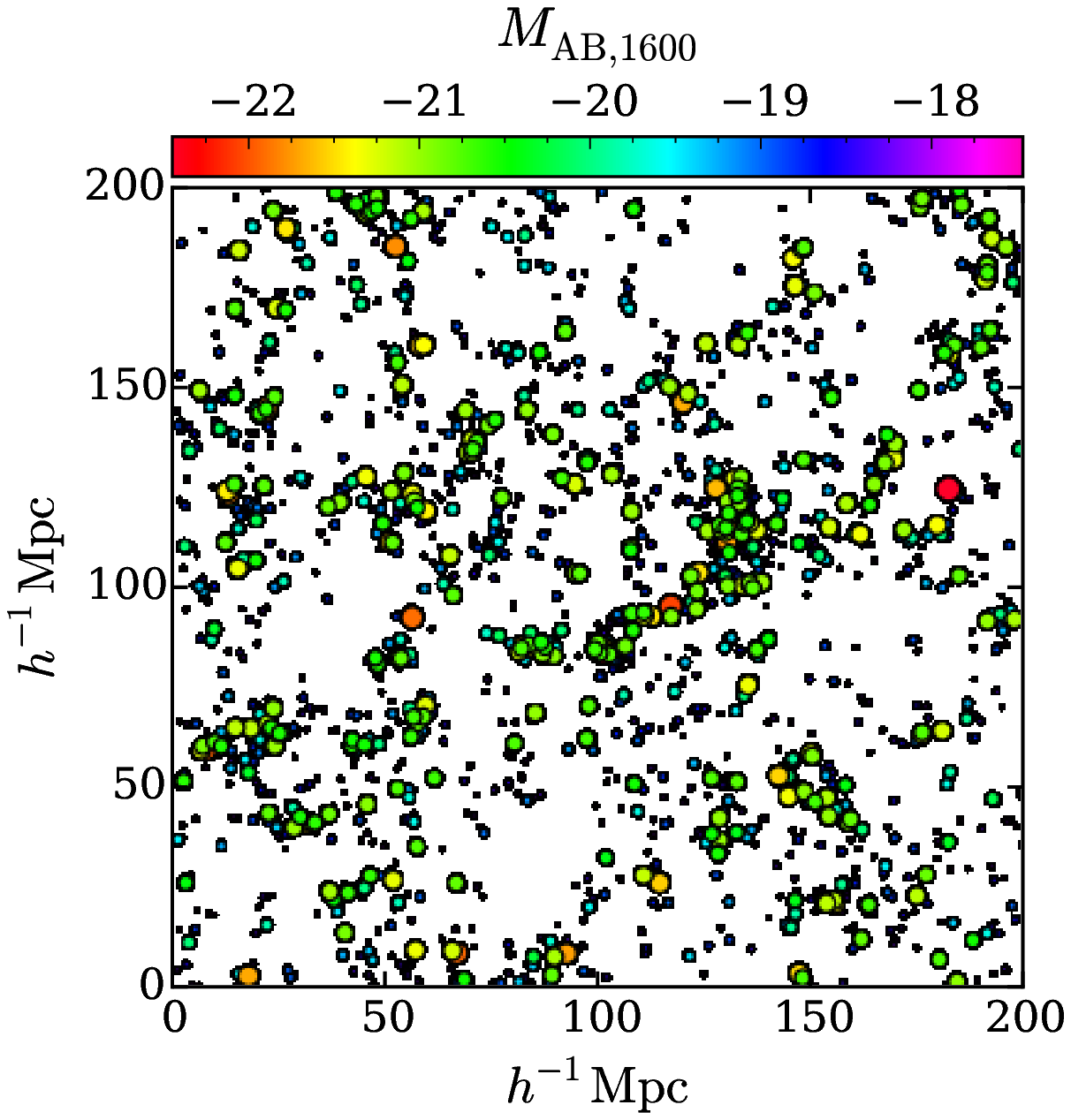}}
\hspace{-0.28cm}
\resizebox{6.0cm}{!}{\includegraphics{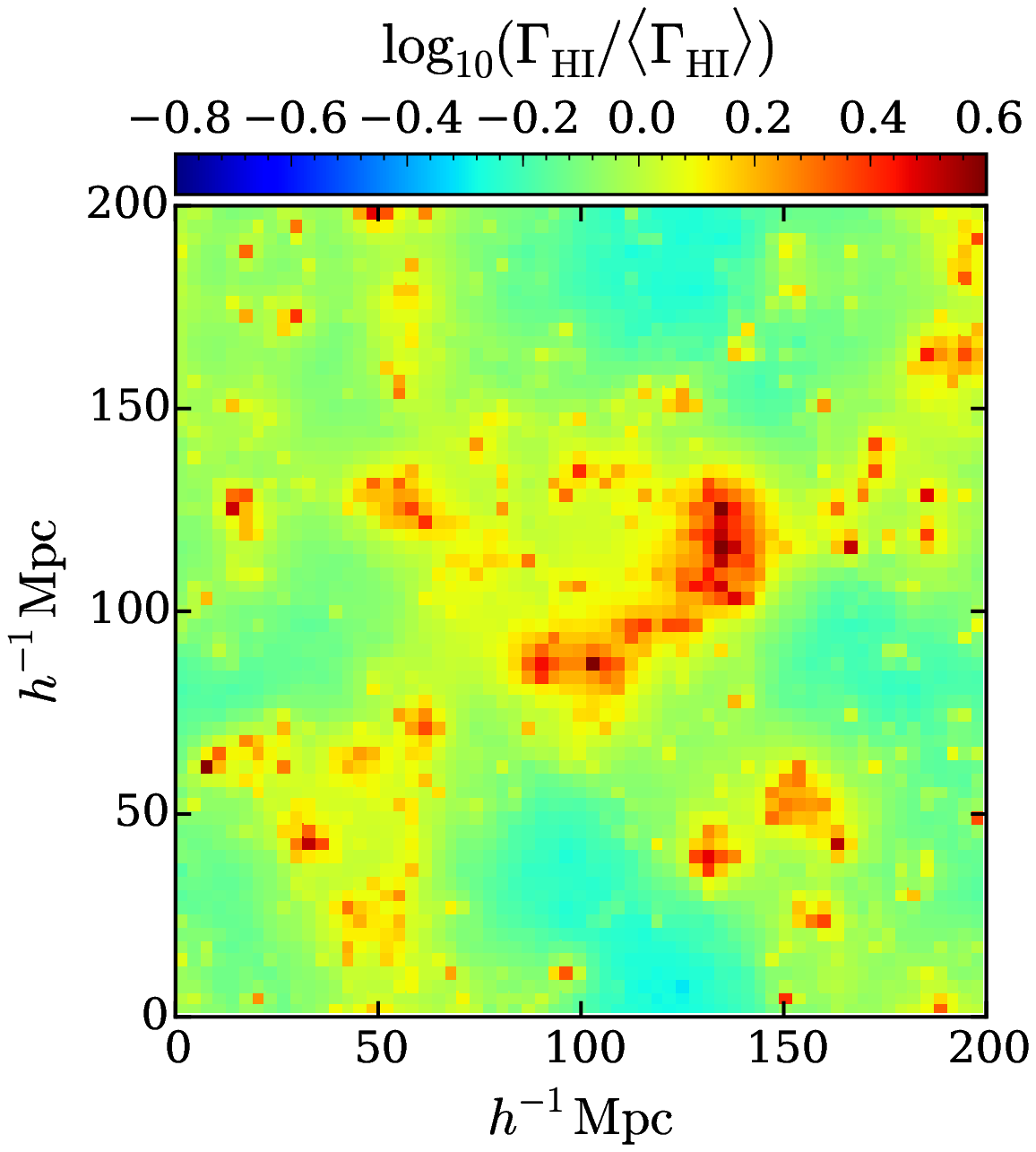}}
\hspace{-0.28cm}
\resizebox{6.0cm}{!}{\includegraphics{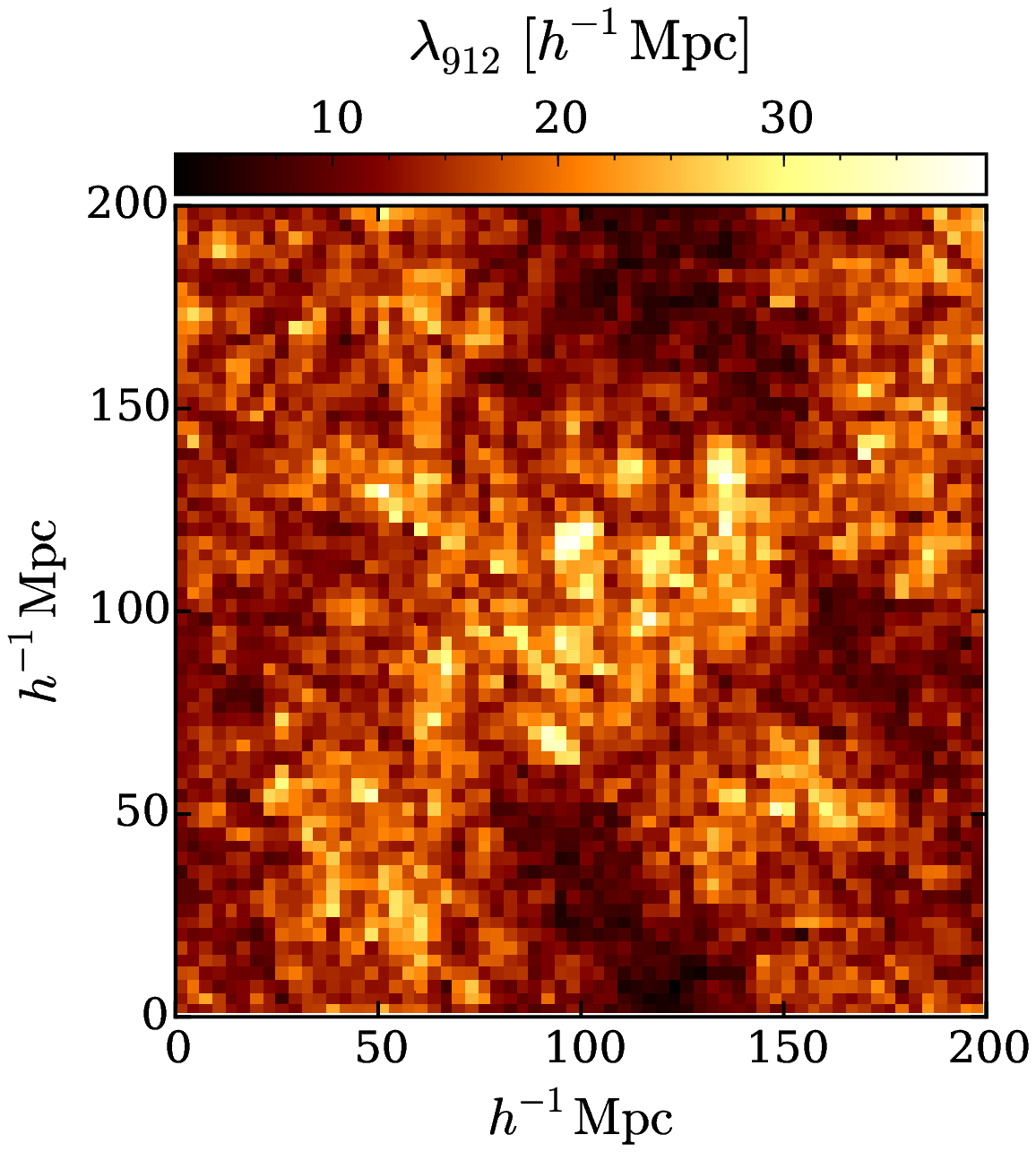}}
\hspace{-0.28cm}
\resizebox{6.0cm}{!}{\includegraphics{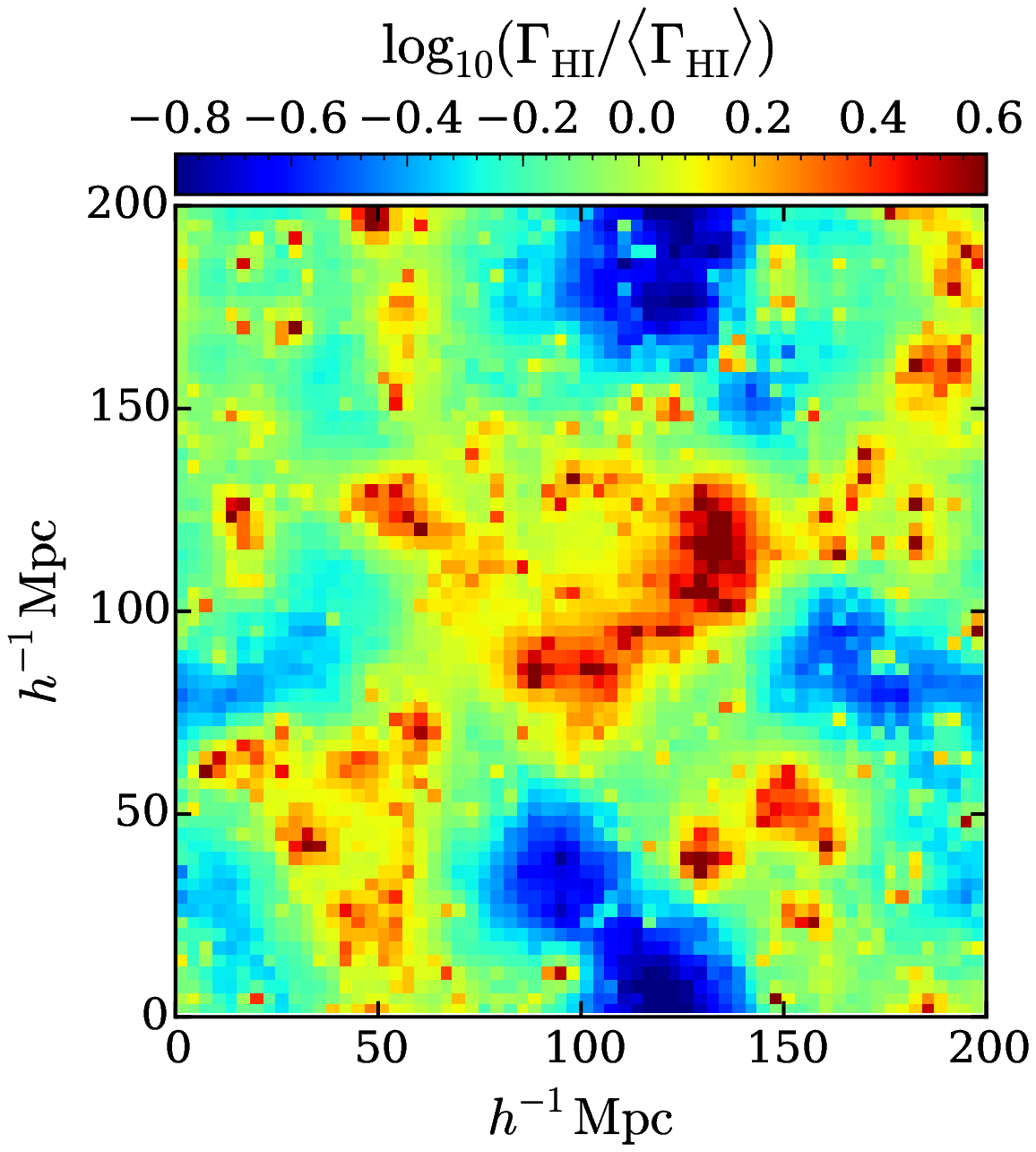}}
\end{center}
\caption{Models of fluctuations in the \HI\ ionizing radiation background at $z=5.6$.  Top left: a slice through the gas density field in our hydro simulation with $L_{\rm box} = 200 h^{-1}~\Mpc$, $N=2048^3$.  The slice thickness is $10 h^{-1}~\Mpc$.  Top middle: galactic sources in the same slice as in the left panel.   The sizes and colors of the points scale with the magnitude of the sources.  For clarity, we have displayed all galaxies in the slice with $M_{\mathrm{AB},1600}<-20.5$, but only a randomly selected 25\% of galaxies with $M_{\mathrm{AB},1600}\geq-20.5$.  Top right: The \HI\ photoionization rate in our fiducial model with $\langle \MFP \rangle = 30 h^{-1}\Mpc$. Here we show a slice of thickness  $3.125h^{-1}~\Mpc$, situated in the middle of the $10h^{-1}~\Mpc$ slice in the left two panels. Bottom left and right: spatial variations in the mean free path in our ``short mean free path" model, with $\langle \MFP \rangle = 15 h^{-1}\Mpc$ (left), and corresponding fluctuations in $\GammaHI$ (right).  The amplitude of $\GammaHI$ fluctuations is much larger in the short mean free path model.     }
\label{FIG:visA}
\end{figure*}

\ctable[ caption = Values of $\langle \MFP \rangle$ for the models in this paper.$^*$]{l c c c c c c}{

\label{TAB:MFPmodels}

\tnote[*]{In units of comoving $ h^{-1}\Mpc$.  }

}{
\hline\hline
Redshifts & $4.8$ & $5$ & $5.2$ & $5.4$ & $5.6$ & $5.8$ \\  \hline
\\ [-1ex]
Fiducial & 54 & 46 & 40 & 35 & 30 & 27 \\
Intermediate & 35 & 30 & 26 & 23 & 20 & 18  \\
Short & 27 & 23 & 20 & 17 & 15 &13  \\
[1.5ex]
\hline
}

We model fluctuations in the ionizing background in post-processing using the approach of \citet{2015arXiv150907131D}.  This approach takes into account spatial variations in $\MFP$ from modulations in the ionization state of optically thick absorbers. We refer the reader to this paper for technical details.  In summary, we solve for the spatially varying $\GammaHI$ field iteratively under the assumption that the mean free path scales as $\MFP(\boldsymbol{x}) \propto \GammaHI^{2/3}(\boldsymbol{x})/\Delta(\boldsymbol{x})$, where $\Delta(\boldsymbol{x})$ is the local matter density.  The $\GammaHI^{2/3}$ scaling is motivated by the analytical model of \citet{2000ApJ...530....1M} (see also \citealt{2005MNRAS.363.1031F}),  and is consistent with the scaling found in the radiative transfer simulations of \citet{2011ApJ...743...82M}.  (In the next section, we will discuss the effect of varying this scaling relation.)  The $\Delta^{-1}$ scaling takes into account the bias of optically thick absorbers with respect to the underlying density field.  We find that our results are insensitive to the particular form of this scaling (see also \citealt{2015arXiv150907131D} and \citealt{2016arXiv160608231C}).    We compute $\GammaHI$ on a uniform grid of $N=64^3$ cells, such that our cell size of $\Delta x=3.125~h^{-1}\Mpc$ is much smaller than $\langle\MFP\rangle$ at all redshifts considered. 
 
We use our models of the fluctuating ionizing background along with our hydro simulation to model the distribution of Ly$\alpha$ forest $\taueff$.  We trace 4000 skewers of length $50h^{-1}~\Mpc$ at random angles through our hydro box, rescaling the \HI\ densities along these skewers according to the $\GammaHI$ values in our background models (assuming photoionization equilibrium).   We then construct synthetic Ly$\alpha$ forest sight lines using the method of \citet{1998MNRAS.301..478T}.  We compare the $\taueff$ fluctuation amplitude in our models to the measurements of \citet{2015MNRAS.447.3402B} by computing the cumulative probability distribution of $\taueff$, $\CPDF$.  At each redshift, we rescale $\GammaHI$ by a constant factor to match the observed mean value of $\langle F \rangle_{50}$.  This procedure fixes the spatially averaged photoionization rate and emissivity, $\langle \GammaHI \rangle$ and $\langle \epsilon_{912} \rangle$, effectively providing a measurement of these quantities (see \S \ref{SEC:globalmeasurements}).  In Appendix \ref{SEC:convergence}, we use a suite of higher resolution test simulations to show that our $\taueff$ distributions are reasonably well converged at the resolution of our hydro simulation.

\begin{figure*}
\begin{center}
\resizebox{18cm}{!}{\includegraphics{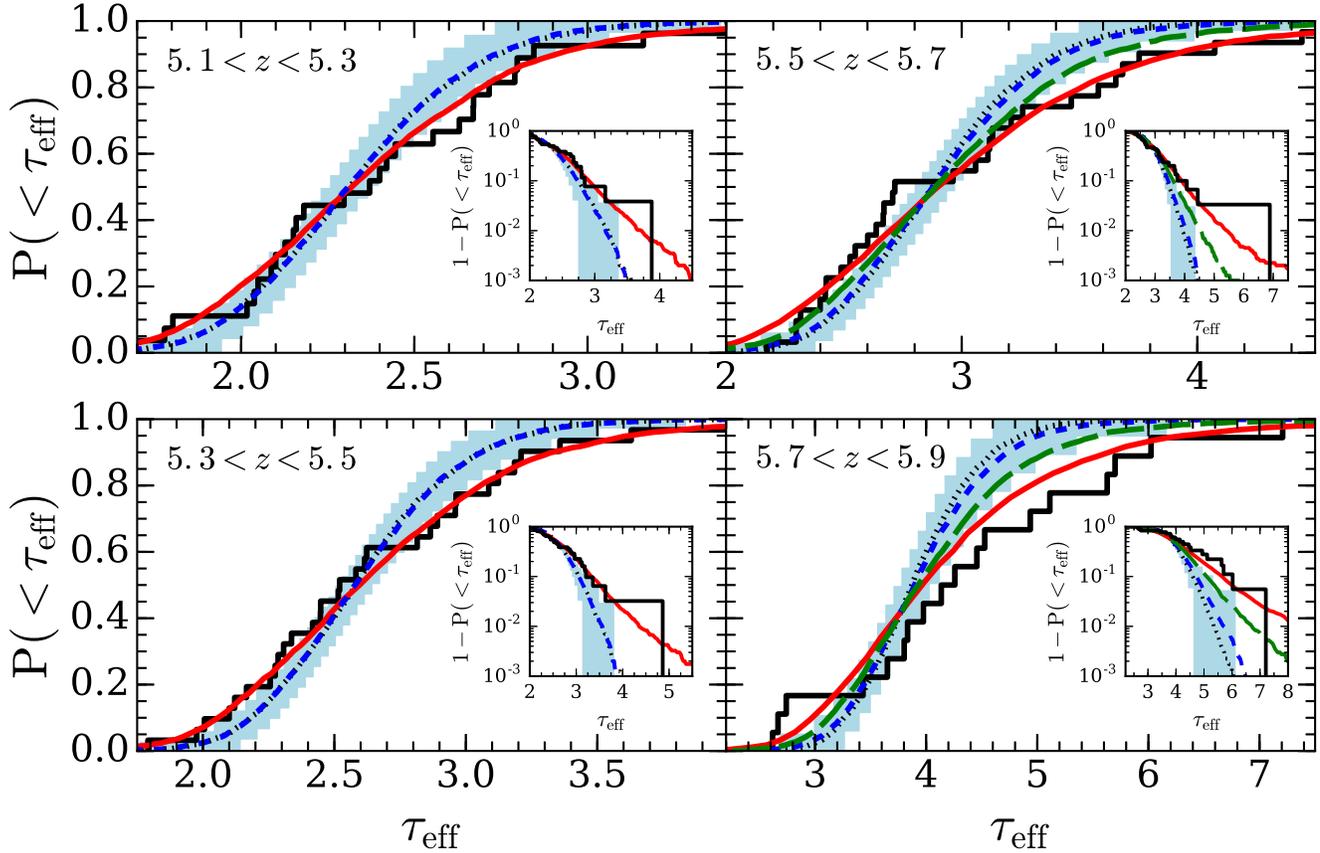}}
\end{center}
\caption{Ly$\alpha$ forest opacity variations in the $z>5$ Ly$\alpha$ forest from fluctuations in the mean free path and ionizing background. The black histograms show the cumulative probability distribution of $\taueff$, $\CPDF$, measured in $50h^{-1}\Mpc$ segments of the forest by \citet{2015MNRAS.447.3402B}. For reference, the black/dotted curves (barely distinguishable from the blue/short-dashed curves) correspond to a model in which the \HI\ photoionization rate is assumed to be uniform. The blue/short-dashed curves show our fiducial model,  consistent with extrapolating recent observational measurements of the mean free path at $z\leq 5.2$ \citep{2014MNRAS.445.1745W}.  The light blue shaded regions show the 90\% confidence limits obtained by bootstrap sampling of this model.  The red curves correspond to our short mean free path model in which $\langle \MFP \rangle$ is a factor of two shorter than in the fiducial model.  In the $z=5.6$ and $z=5.8$ bins, the green/long-dashed curves correspond to our intermediate scenario, which interpolates between the fiducial and short models.  The insets show $1-\CPDF$ on a logarithmic scale, providing a detailed view of the high-opacity tail of $\CPDF$. }
\label{FIG:taueff_galaxies}
\end{figure*}

 \subsection{Models}  
 
 In what follows, we will consider three models for the mean free path.  The values of $\langle \MFP \rangle$ are summarized in Table \ref{TAB:MFPmodels}.  Fig. \ref{FIG:MFPmodels} compares these models against a compilation of recent observational measurements. For reference, the black/dotted curve shows the empirical fit of \citet{2014MNRAS.445.1745W}\footnote{{\addition Equation (\ref{EQ:MFP}) is equivalent to the fit given in \citet{2014MNRAS.445.1745W} when cast in comoving units and in terms of $h=H_0/(100 \mathrm{km}/\mathrm{s}/\Mpc)$.}}, 
\begin{equation} 
\MFP = 130~h^{-1}\mathrm{Mpc} \left( \frac{1+z}{5}\right)^{-4.4}.
\label{EQ:MFP}
\end{equation}  
In our ``fiducial" model, $\langle \MFP \rangle(z)$ is consistent with the extrapolation of the measurements in \citet{2014MNRAS.445.1745W} to higher redshifts. In our ``short mean free path" model, $\langle \MFP \rangle$ is a factor of two lower than the fiducial case, while the ``intermediate" model falls halfway between these models. The top left, middle, and right panels of Fig. \ref{FIG:visA} show slices through the gas density, galaxy distribution, and photoionization rate, respectively, in our fiducial model at $z=5.6$.  The left and middle slices are of thicknesses $10 h^{-1}~\Mpc$, while the $\GammaHI$ slice is of thickness $3.125h^{-1}~\Mpc$, situated in the middle of the $10h^{-1}~\Mpc$ slice.  In the middle panel the sizes and colors of the points scale with the magnitude of the galaxy.  For clarity, we have displayed all galaxies in the slice with $M_{\mathrm{AB},1600}<-20.5$, but only a randomly selected 25\% of galaxies with $M_{\mathrm{AB},1600}\geq-20.5$.   In the bottom row, we show a slice through the mean free path field (left panel) and the corresponding fluctuations in $\GammaHI$ (right panel) for our short mean free path model.   In this case, the fluctuations in $\GammaHI$ are much larger.  
  
\subsection{Results}
\label{SEC:opacityflucs}

Fig. \ref{FIG:taueff_galaxies} compares $\CPDF$ in our models to the measurements of \citet{2015MNRAS.447.3402B} (black histograms).  The blue/short-dashed curves show our fiducial model.  The light blue shaded regions show the 90\% confidence limits obtained by bootstrap sampling from the model the same number of sight lines in each redshift bin as the observations, where we fix the mean $\langle F \rangle_{50}$ of each sample to the observed value. For reference, the black dotted curves, which are barely distinguishable from the blue/short-dashed curves, correspond to a model in which $\GammaHI$ is assumed to be uniform.  The red/solid curves show the short mean free path model in which $\langle \MFP \rangle$ is a factor of two shorter.   In the $z=5.6$ and $5.8$ bins, the green/long-dashed curves correspond to the intermediate scenario.  The insets in Fig. \ref{FIG:taueff_galaxies} show $1-\CPDF$ on a logarithmic scale, providing a detailed view of the high-opacity tail of $\CPDF$. 

\begin{figure}
\begin{center}
\resizebox{8.8cm}{!}{\includegraphics{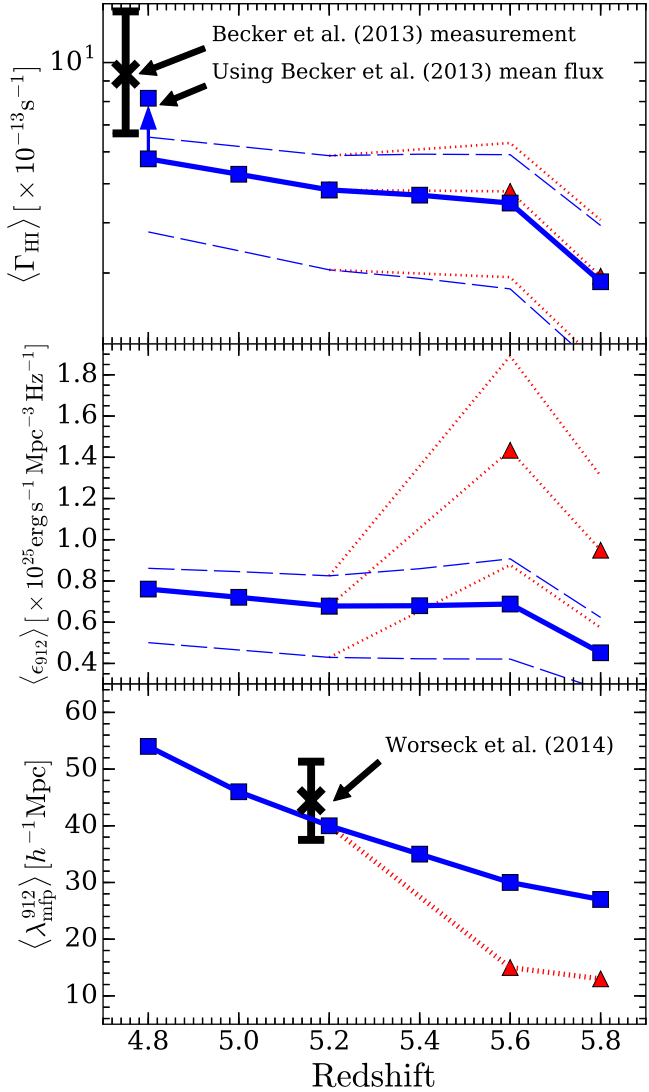}}
\end{center}
\caption{Measurements of the spatially averaged \HI\ photoionization rate (top) and of the ionizing emissivity (middle), {\addition for models assuming the mean free path values in the bottom panel.}  The blue squares correspond to our fiducial model for $\langle \MFP \rangle$, consistent with observational measurements of the mean free path at $z\lesssim 5.2$ (see Table \ref{TAB:MFPmodels} and Fig. \ref{FIG:MFPmodels}).  The thin dashed curves represent uncertainties in the thermal history of the IGM.   For the red triangles, we assume our short mean free path model at $z=5.6$ and $z=5.8$, i.e. the values of $\langle \MFP \rangle$ favored by the observed amplitude of Ly$\alpha$ forest opacity fluctuations (here the thermal history uncertainties are represented by dotted curves for clarity).  This plot shows that the ionizing emissivity of the galaxy population is required to evolve by a factor of $\approx 2$ in the $\approx 100$ million years between $z=5.2-5.6$ to be simultaneously consistent with the observed opacity fluctuations at $z=5.6$ and the mean free path measurements at $z\leq 5.2$.}   
\label{FIG:galaxy_em}
\end{figure}

Let us begin by considering the models with uniform $\GammaHI$ (black/dotted). These models are marginally consistent with most of the observed distribution at $z=5.2$ and at $z=5.4$.\footnote{Our comparison of the observed $\CPDF$ to our models differs from the comparisons in \citet{2015MNRAS.447.3402B} in an important way. \citet{2015MNRAS.447.3402B} rescaled $\GammaHI$ such that their model distributions are equal to the observed distribution at a fixed point at low $\taueff$ (see e.g. their Fig. 11). In contrast, we rescale $\GammaHI$ to match the mean $\langle F \rangle_{50}$ to the observed value.  Not only does our procedure yield a higher mean $\taueff$; it also yields a somewhat broader $\CPDF$ because of the nonlinear relationship, $\taueff = - \ln \langle F \rangle_{50}$. Thus, normalizing to the mean $\langle F \rangle_{50}$ alleviates some of the tension between the uniform $\GammaHI$ model and observations, especially at $z\lesssim 5.5$.  Fitting the observed $\taueff$ distribution with the free parameter $\langle \GammaHI \rangle$ may further alleviate the tension.} They do, however, fail to account for the highest opacity measurements at those redshifts.  The discrepancy with observations clearly grows with redshift as shown in the $z=5.6$ and $z=5.8$ bins.  Comparing the blue/short-dashed and black/dotted curves in Fig. \ref{FIG:taueff_galaxies} shows that $\CPDF$ in our fiducial model is very similar to that in the uniform $\GammaHI$ model. This indicates that background fluctuations have a minimal impact on the $\taueff$ distribution if $\langle \MFP \rangle$ is similar to expectations from measurements at $z\lesssim 5.2$.  In Appendix \ref{SEC:DLAs}, we extend our models to include Damped Ly$\alpha$ systems (DLAs).  There we show that this conclusion is unaffected by the presence of DLAs under reasonable assumptions about their abundance.  On the other hand, our short mean free path model is able to reproduce the full width of the observed $\taueff$ distribution.  This is qualitatively consistent with the results of \citet{2015arXiv150907131D}, except that even smaller values ($\langle \MFP \rangle \approx 10h^{-1}~\Mpc$ at $z=5.6$) were favored -- a difference that likely owes to their lower minimum halo mass of $\Mmin = 2\times10^{9}~\Msun$ for hosting galactic sources.  Lower values of $\Mmin$ results in smaller fluctuations because less massive halos are more weakly clustered. {\addition In this sense, our results can be viewed as an upper limit to the $\langle \MFP \rangle$ values required to reproduce the opacity measurements of \citet{2015MNRAS.447.3402B}.}  Finally, we note that our conclusions are relatively insensitive to the scaling of the local $\MFP$ with $\GammaHI$.  We find that models with $\MFP \propto \GammaHI^{0.75}$, i.e. the steepest scaling found in the radiative transfer simulations of \citet{2011ApJ...743...82M}, yield $\taueff$ distributions that are only $1-2$ line widths wider than those in Fig. \ref{FIG:taueff_galaxies} (which assume $\MFP \propto \GammaHI^{2/3}$).

\section{The global photoionization rate and ionizing emissivity}
\label{SEC:globalmeasurements}

\ctable[ caption = Measurements of the global \HI\ photoionization rate and the ionizing emissivity at $z\approx 4.8 - 5.8$, star]{c c c c c c c}{

\label{TAB:GammaHI}

\tnote[*]{From the mean values of $\langle F \rangle_{50}$ in \citet{2015MNRAS.447.3402B}.  Here we show only statistical uncertainties estimated by bootstrap sampling.}
\tnote[$\dagger$]{In units of $10^{-12}$ s$^{-1}$.  The upper and lower limits on all values of $\langle \GammaHI \rangle$ and $\langle \epsilon_{912} \rangle$ generously bracket uncertainties in the thermal state of the IGM (see Appendix \ref{SEC:sim_thermal_hist}).  } 
\tnote[$\ddagger$]{In units of $\times10^{25}$ erg s$^{-1}$ Mpc$^{-3}$ Hz$^{-1}$.}

}{
\hline\hline
Redshifts & $4.7 < z < 4.9$ & $4.9 < z < 5.1$ & $5.1 < z < 5.3$ & $5.3 < z < 5.5$ & $5.5 < z < 5.7$ & $5.7 < z < 5.9$ \\  \hline
\\ [-1ex]
Mean transmission$^*$ & $0.16\pm0.01$ & $0.14\pm0.01$ & $0.10\pm0.01$ & $0.080\pm0.006$ & $0.06\pm0.005$ & $0.022\pm0.005$ \\ [1.5ex]
\hline 
 Fiducial & & & & & & \\
\hline \\ [-1.5ex]
$\langle \GammaHI \rangle^\dagger$ & $0.58^{+0.08}_{-0.20}$ & $0.53^{+0.09}_{-0.19}$ & $0.48^{+0.10}_{-0.18}$ & $0.47^{+0.12}_{-0.18}$ & $0.45^{+0.14}_{-0.17}$ & $0.29^{+0.11}_{-0.11}$  \\  [1.5ex]
$\langle \epsilon_{912} \rangle^\ddagger$ & $0.76^{+0.10}_{-0.26}$ & $0.72^{+0.12}_{-0.26}$ & $0.68^{+0.15}_{-0.25}$ & $0.68^{+0.18}_{-0.26}$ & $0.69^{+0.22}_{-0.27}$ & $0.45^{+0.17}_{-0.18}$  \\ [1.5ex]
\hline 
 Short Mean Free Path & & & & & & \\
\hline \\ [-1.5ex]
$\langle \GammaHI \rangle$ & $0.60^{+0.08}_{-0.20}$ & $0.55^{+0.10}_{-0.20}$ & $0.51^{+0.11}_{-0.19}$ & $0.50^{+0.13}_{-0.19}$ & $0.48^{+0.15}_{-0.19}$ & $0.29^{+0.11}_{-0.12}$  \\ [1.5ex]
$\langle \epsilon_{912} \rangle$ & $1.31^{+0.17}_{-0.45}$ & $1.31^{+0.23}_{-0.46}$ & $1.29^{+0.28}_{-0.48}$ & $1.39^{+0.37}_{-0.53}$ & $1.43^{+0.46}_{-0.56}$ & $0.95^{+0.36}_{-0.38}$  \\
[1.5ex]
\hline
}

Rescaling $\langle \GammaHI \rangle$ in our models to match the observed mean transmission in the forest effectively yields a measurement of $\langle \GammaHI\rangle$ and of the global ionizing emissivity, $\langle \epsilon_{912} \rangle$.  In this section we present new measurements of these quantities at $z=4.8-5.8$ based on the $\taueff$ measurements of \citet{2015MNRAS.447.3402B}.  Whereas previous measurements at lower redshifts have assumed a uniform ionizing background, ours account for fluctuations in the background and in the mean free path. {\addition For comparison, we also performed our measurements assuming a uniform $\GammaHI$. We find that it leads to $\langle \GammaHI \rangle$ that are only $5-10\%$ lower than our fiducial measurements including fluctuations.  This is consistent with the finding of \citet{2009MNRAS.400.1461M} that fluctuations in $\GammaHI$ have only a mild impact on Ly$\alpha$ forest measurements of $\langle \GammaHI \rangle$.}   Table \ref{TAB:GammaHI} reports the measurements from our fiducial and short mean free path models.\footnote{Our measurements of $\langle \epsilon_{912} \rangle$ include a crude correction to account for the contribution of recombination radiation to the ionizing background. Assuming an IGM temperature of $T=10,000$ K, $(\alpha_{\rm A} - \alpha_{\rm B})/\alpha_{\rm A} \approx 40\%$ of radiative recombinations produce an \HI\ ionizing photon, where $\alpha_{\rm A}$ and $\alpha_{\rm B}$ are the case A and case B recombination coefficients, respectively.  Only a fraction of these photons escape the optically thick systems in which they are produced.  Under standard assumptions about the distribution of optically thick absorbers, the fraction of photons that escape is approximately one half.  We therefore scale down our emissivities by $20 \%$.  We note that this crude estimate is consistent with more detailed studies, which have obtained a correction of $\approx 10-30 \%$ at these redshifts \citep{2009ApJ...703.1416F}.}  The top row gives our estimates of the mean transmission in the forest, obtained from the mean values of $\langle F \rangle_{50}$ in \citet{2015MNRAS.447.3402B}.   

The blue squares in Fig. \ref{FIG:galaxy_em} show the measurements of $\langle \GammaHI \rangle$ (top) and $\langle \epsilon_{912}\rangle$ (middle) from our fiducial model.  {\addition Our measurements include a correction to account for the effects of finite simulation resolution.  These corrections are described in Appendix \ref{SEC:sim_thermal_hist}.}  The thin dashed curves bracketing these data points correspond to generous upper and lower limits on the effect of the IGM thermal state, also described in Appendix \ref{SEC:sim_thermal_hist}.   Although these limits do not represent the full range of uncertainties in our measurements, we note that they are the dominant modeling uncertainties.   For comparison, the black ``x" in the top panel shows the highest redshift measurement of \citet{2013MNRAS.436.1023B}. We find a somewhat lower value for $\langle \GammaHI \rangle$ at $z=4.8$.  The difference is mostly due to an offset in the mean flux between the data of \citet{2013MNRAS.436.1023B} and \citet{2015MNRAS.447.3402B}, where we have used the latter for our measurement. The vertical arrow at $z=4.8$ shows the higher value of $\langle \GammaHI \rangle$ that is obtained by normalizing our simulations to the mean flux in \citet{2013MNRAS.436.1023B}.  The red triangles at $z=5.6$ and $5.8$ correspond to our short mean free path model (recall that this model better accounts for the observed dispersion in $\taueff$ at those redshifts). For clarity we depict the thermal state uncertainties for this model as thin dotted lines.  {\addition We note that our measurements of $\langle \GammaHI \rangle$ are consistent with the proximity region measurements of \citet{2011MNRAS.412.2543C} and \citet{2011MNRAS.412.1926W}.  For example, the former report $\log(\GammaHI/[\mathrm{s}^{-1}]) = -12.15\pm0.16$ and $-12.84\pm0.18$ ($1\sigma$) at $z\approx5$ and $6$, respectively.  }  

The global photoionization rate stays remarkably flat over the interval $z\approx 4.8-5.6$, before decreasing sharply at $z\approx 5.8$. Our measurements of the corresponding emissivities reveal a problematic feature of the short mean free path model.  Because $\langle \GammaHI \rangle(z)$ is so flat for $z\lesssim 5.6$, increasing $\langle \MFP \rangle$ by a large factor between $z=5.6$ and $z=5.2$ (where the mean free path has been measured to be $\approx 44~h^{-1}\Mpc$ ) requires a decrease in $\langle \epsilon_{\rm 912}\rangle$ by a similarly large factor.  This effect is shown in the middle panel of Fig. \ref{FIG:galaxy_em}.  Following the central red/dotted curve, we find that $\langle \epsilon_{912} \rangle$ has to evolve by a factor of $\approx 2$ over the $\approx 100$ million years between $z=5.2-5.6$.  This time interval is substantially shorter than the Hubble time of a billion years, which is (within a factor of a few) the time scale over which we should expect such a large change in the output of the galaxy population. Note that the required evolution in $\langle \epsilon_{912} \rangle$ would only be steeper if our simulations included the less biased galaxies that have luminosities below current detection limits, i.e. if $\Mmin$ were lower than our assumed value of $2\times10^{10}h^{-1}~\Msun$.   It is unlikely that feedback on dwarf galaxies from the photoheating of the IGM by reionization can induce such a rapid evolution in the ionizing emissivity, as simulations show that this feedback is more gradual \citep[e.g.][]{2000ApJ...542..535G,2004ApJ...601..666D, 2006MNRAS.371..401H, 2008MNRAS.390..920O, 2013MNRAS.432L..51S, 2014MNRAS.444..503N}. {\addition Lastly, we note that the rapid jump in the emissivity at $z=5.8$ is an artifact of the fact that we have adopted the redshift scaling of equation (\ref{EQ:MFP}) for $\langle \MFP \rangle$.  The rapid evolution in $\langle \GammaHI \rangle$ may indicate that $\langle \MFP \rangle$ is evolving more steeply with redshift towards $z\sim 5.8$. } 
 
In these models, there are two ways to avoid the problem of a rapidly evolving $\langle \epsilon_{912} \rangle$ while at the same time accounting for the large dispersion in $\taueff$ at $z\gtrsim5.6$:  

\medskip
\noindent (i) Fluctuations in the ionizing background are driven by sources that are much rarer and brighter than faint, sub-$L*$ galaxies. Such a scenario would remove the need for a short mean free path.  

\medskip
\noindent
(ii) The direct measurement of the mean free path at $z \approx 5.2$ by \citet{2014MNRAS.445.1745W} does not probe $\langle \MFP \rangle$, but is instead biased substantially higher by quasar line-of-sight effects. 

\medskip
\noindent
In Appendix \ref{SEC:fesc}, we explore models in which the escape fraction increases with UV luminosity, such that the ionizing emissivity is weighted towards rarer, brighter galaxies than in our fiducial model.  We show that accounting for the observed $\taueff$ fluctuations at $z=5.6$ requires that $f_{\rm esc}\gtrsim 50\%$ for galaxies with $M_{\rm AB, 1600} \lesssim M^*_{\rm AB, 1600} \approx -21$ (corresponding to $M_{200} \gtrsim 10^{11} h^{-1}~\Msun$), and $f_{\rm esc} \sim 0$ for fainter galaxies.\footnote{ In our fiducial model, the escape fraction is already required to be $f_{\rm esc} \gtrsim 10$ \%.}  In this scenario, the ionizing emissivity is driven by galaxies with space density $\lesssim 10^{-4}~\Mpc^{-3}$.  We note that this would differ from the situation at $z\approx 3$, where observations indicate that $L > 0.5 L^*$ galaxies with large $f_{\rm esc}$ are extremely rare \citep[e.g.][]{2010ApJ...725.1011V, 2015ApJ...810..107M, 2015ApJ...804...17S, 2016A&amp;A...585A..48G}.   Alternatively, some authors have proposed models in which AGN contribute significantly to the $z>5$ \HI\ ionizing background \citep{2015arXiv150501853C, 2016arXiv160608231C}.   However, in Paper II we argue that these models are in tension with recent constraints on the thermal history of the IGM.  In the next section, we will explore the possibility that measurements of the mean free path are biased by quasar line-of-sight effects.   

\section{The effect of quasar proximity zones on measurements of the mean free path}
 \label{SEC:MFPbias}

The arguments presented in the last section rely on the assumption that the measurement of $\MFP(z\approx5.2) = 44 \pm 7~h^{-1}\Mpc$ by \citet{2014MNRAS.445.1745W} is equivalent to a measurement of $\langle \MFP \rangle$.  In this section we will show that the measurement may in fact be biased significantly higher than $\langle \MFP \rangle$ owing to the enhanced ionizing flux along quasar sight lines.

The measurements of \citet{2014MNRAS.445.1745W} were obtained by stacking rest-frame quasar spectra and fitting a simple model (under the assumption of a uniform ionizing background) for the mean transmission blue-ward of the Lyman limit.  These authors point out that such measurements could be affected by the enhanced transmission in the proximity zones of the quasars in the stack, particularly at high redshifts. \citet{2014MNRAS.445.1745W} concluded that this effect is likely negligible for their $z\approx 5.2$ measurement in part because it is consistent with an extrapolation of measurements at $z\lesssim 5$ (which, they argue, are less affected by proximity zones).  

However, at $z\approx 5.2$, we find that the proximity zones can extend over spatial scales that are comparable to or larger than $\langle \MFP \rangle$.  As a simple illustration of this, we compute the radius $R$ at which the contribution to $\GammaHI$ from a quasar with specific intensity $J_{\nu} \propto 1/R^2$ is equal to the background value, $\GammaHI^{\rm bkgd}$.  For a $z=5.2$ quasar\footnote{Here and throughout this section, we assume a quasar spectrum with specific luminosity $L_{\nu} \propto\nu^{-0.6}$ at $\lambda >912$ \AA, and $L_{\nu} \propto \nu^{-1.7}$ for $\lambda \leq 912$ \AA, consistent with the stacked quasar spectrum of \citet{2015MNRAS.449.4204L}} with $M_{\rm AB,1450}=-27$, corresponding to roughly the mean luminosity of the $5 < z < 5.5$ sample in \citealt{2014MNRAS.445.1745W}, we find $R \approx 81, 36$ and $26$ (comoving) $h^{-1}\Mpc$ for $\GammaHI^{\rm bkgd}=0.1, 0.5,$ and $1\times10^{-12}\mathrm{s}^{-1}$, respectively.  We further note that the proximity zone need not extent out to $R=\langle \MFP \rangle$ to have a significant impact on measurements of the {\it integrated} Lyman limit opacity along sight lines emanating from the quasar.  In addition, the quasars likely reside in highly biased halos, such that the local ionizing background is further enhanced by the overdensity of neighboring sources, making the mean free path longer than in an average region.   

\begin{figure}
\begin{center}
\resizebox{8.4cm}{!}{\includegraphics{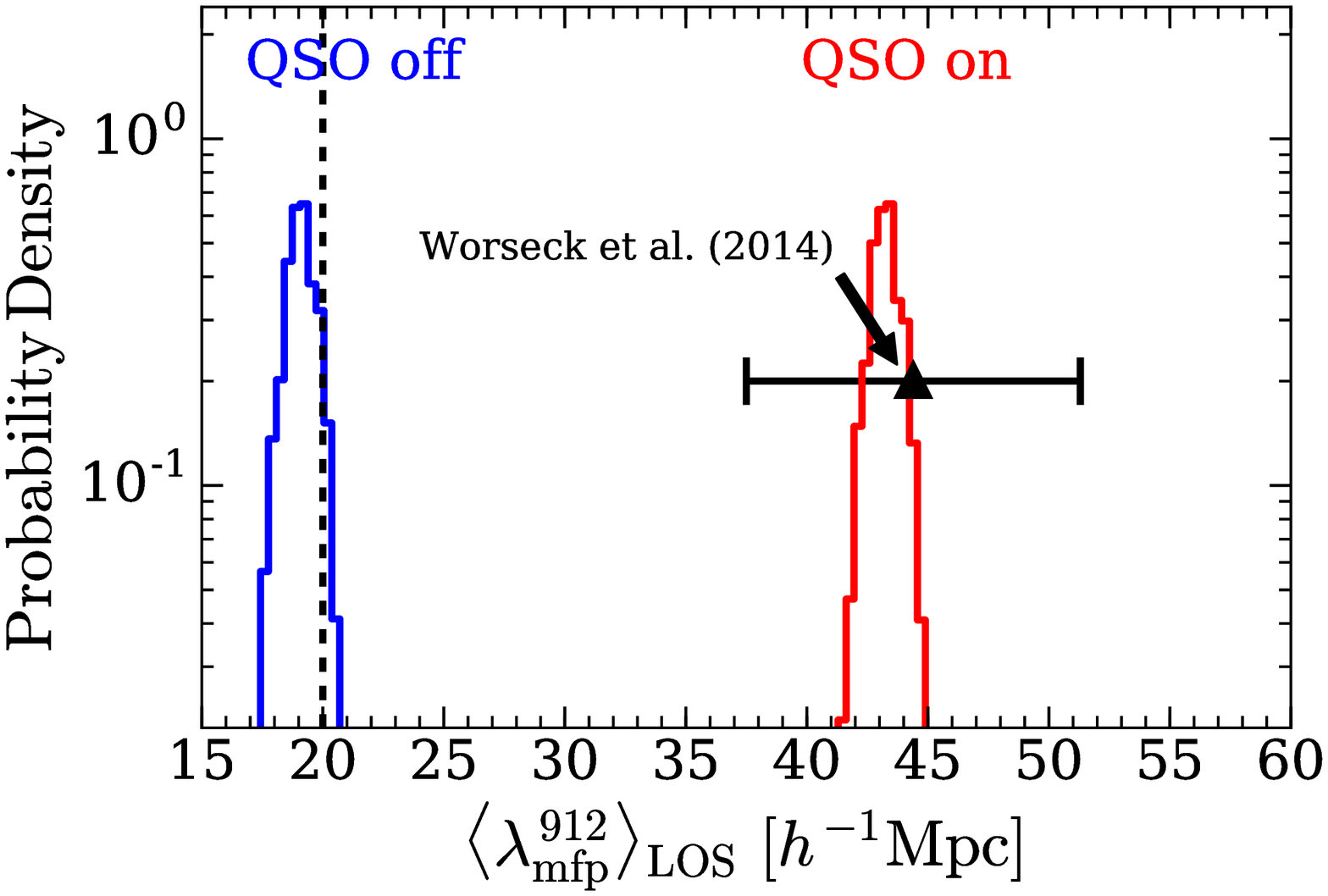}}
\resizebox{8.4cm}{!}{\includegraphics{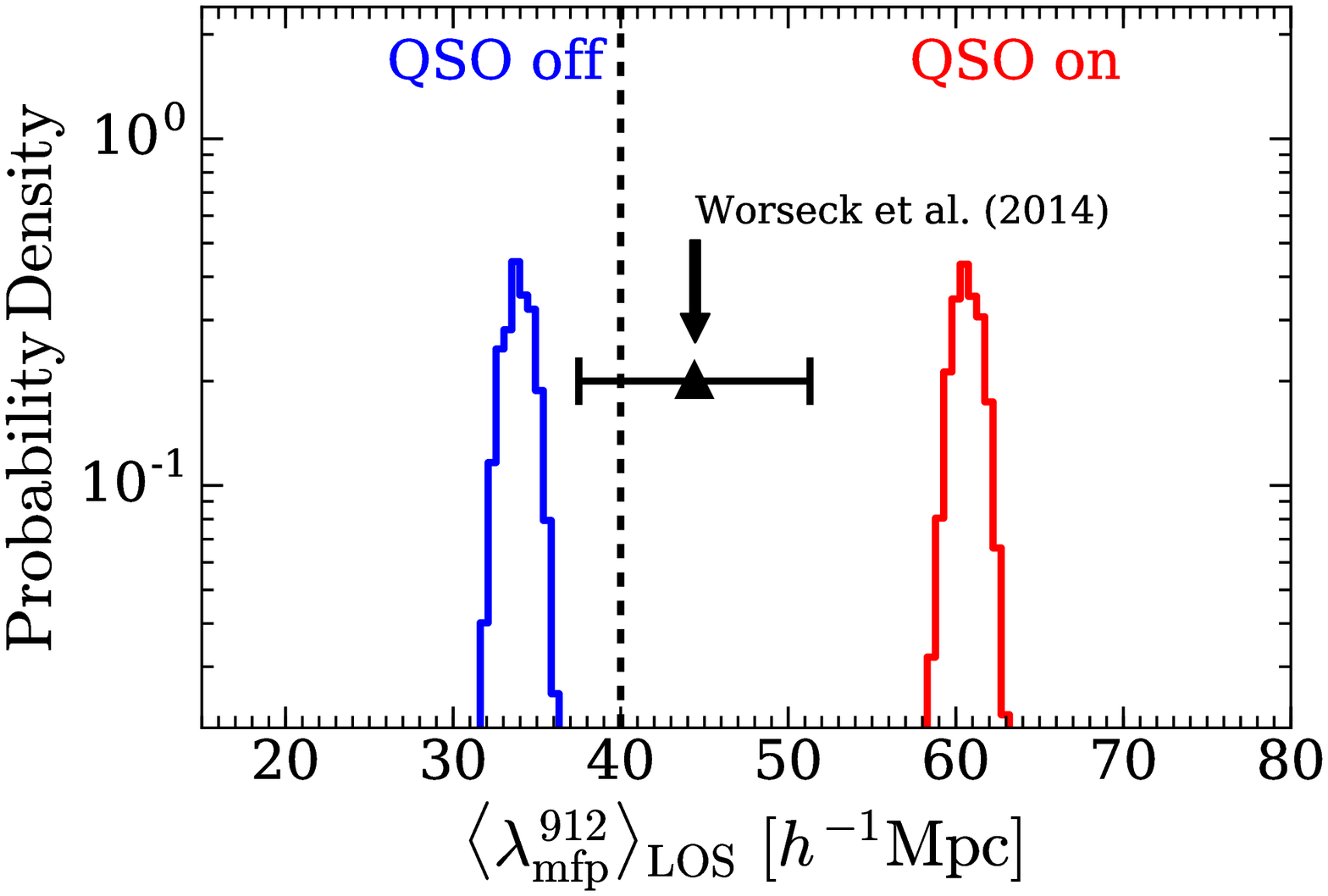}}
\end{center}
\caption{Probability distribution function for measurement of the mean free path from stacked quasar spectra.  We use simulations of the ionizing background around the $39$ quasars in the $5 < z < 5.5$ sample in \citet{2014MNRAS.445.1745W}.  The quantity, $\langle \MFP \rangle_{\rm LOS}$, represents the ``measured" value, i.e. the mean distance to optical depth unity, averaged over the ensemble of $39$ quasars.  Top panel: We adopt our short mean free path model, with $\langle \MFP \rangle=20~h^{-1}\Mpc$, as shown by the vertical dashed line. The enhanced ionizing flux along quasar sight lines leads to the measured value being a factor of $\approx2$ longer than $\langle \MFP \rangle$.  This plot shows that $\langle \MFP \rangle$ might be substantially shorter at $z>5$ than current measurements suggest.   {\addition Bottom panel: Same as the top, except for $\langle \MFP \rangle=40~h^{-1}\Mpc$.  In this case, the bias is reduced to $\approx 50 \%$, and the enhanced densities around the massive halos result in the quasars-off value being $\approx 15 \%$ lower than $\langle \MFP \rangle$.}      }  
\label{FIG:MFPbias}
\end{figure}

Here we quantify the impact of these effects on measurements of the mean free path by simulating the ionizing background around the $39$ quasars in the $5 < z < 5.5$ sample of \citealt{2014MNRAS.445.1745W}.  We perform one simulation for each quasar by assigning its luminosity to one of the $39$ most massive halos in our hydro simulation box.  {\addition Note that, for each simulation, the quasar is assigned to a different halo to allow for environmental variations.  For reference, the quasars in the sample span the range of magnitudes, $M_{\rm AB,1450}= -25.0$ to $-28.2$.} As before, we iteratively solve for $\GammaHI$ around the quasar using the procedure of \citet{2015arXiv150907131D}.  This procedure approximates the Lyman continuum absorption along a sightline as a smooth function of distance.  In reality, the mean free path along any given sightline likely depends more on the incidence of an optically thick absorber.  However, the smooth approximation is sufficient for our purposes given that we are ultimately concerned with the mean free path from {\it stacked} quasar spectra. In all of the simulations we set $\langle \GammaHI \rangle  = 4.8\times10^{-13}~\mathrm{s}^{-1}$, consistent with the value measured at $z=5.2$ (see \S \ref{SEC:globalmeasurements}), and $\langle \MFP \rangle = 20 h^{-1}\Mpc$.  The latter corresponds to the short mean free path model from \S \ref{SEC:opacityflucs}.  We shoot $10,000$ randomly oriented sight lines from each quasar, measuring the distance to a Lyman limit optical depth of unity, $R(\tau_{912} = 1)$, along each sightline.   We then select one sight line from each quasar and compute the average $R(\tau_{912} = 1)$ over the sample, which we denote by $\langle \MFP \rangle_{\rm LOS}$.  This average value represents the mean free path that is measured from the quasar stack.  Note that this procedure yields $10,000$ realizations of the measurement.   

The red histogram in the top panel of Fig. \ref{FIG:MFPbias} shows the distribution of $\langle \MFP \rangle_{\rm LOS}$, while the blue histogram on the left shows the case in which the quasars are switched off.  The vertical dashed line shows the input value, $\langle \MFP \rangle = 20~h^{-1}\Mpc$.  We find that $\langle \MFP \rangle_{\rm LOS} \approx 43~h^{-1}\Mpc$, more than a factor of $2$ longer than the input value.  Curiously, this biased value is consistent with the measurement of $44\pm 7~h^{-1}\Mpc$ by  \citet{2014MNRAS.445.1745W}.  We find that the bias comes primarily from the enhanced ionizing flux contributed by the quasar.  To illustrate this, we add the quasars to our ``quasars-off" simulation in post-processing, such that we do not model the ``back-reaction" of the quasars on the contribution to $\GammaHI$ from the local galaxy population.  In this case, we find that the mean value of $\langle \MFP \rangle_{\rm LOS}$ is $37~h^{-1}\Mpc$. This is only $\approx15\%$ lower than in our ``quasars on" simulations.  {\addition To explore how the bias depends on the value of $\langle \MFP \rangle$, we have also tested a case where $\langle \MFP \rangle = 40h^{-1}~\Mpc$, consistent with our fiducial model (bottom panel of Fig. \ref{FIG:MFPbias}).  In this case, we find that the bias between the ``measured" value and $\langle \MFP \rangle$ is reduced to $\approx 50\%$.  Note also that, in this case, the enhanced densities around the massive halos in our simulation result in the quasars-off value being $\approx 15 \%$ lower than $\langle \MFP \rangle$.   }

From these results we are led to conclude that $\langle \MFP \rangle(z=5.2)$ could be up to a factor of $\approx 2$ shorter than indicated by direct measurements.  In this case, $\langle \MFP \rangle$ is similar to the values required in our short mean free path model to account for the large dispersion in $\taueff$.  If confirmed, this scenario has implications for the ionizing photon budget during reionization.  Adopting values of $\langle \MFP \rangle$ that are a factor of two shorter than were assumed in previous studies results in ionizing emissivities that are a factor of two larger.  To illustrate this, we compute $\dot{N}_{\rm ion} \equiv \langle n_{\rm H}\rangle^{-1} \int^\infty_{\nu_{912}} \dd \nu \epsilon_{\nu}/ (h_{\rm P} \nu$), where $\langle n_H \rangle$ is the mean hydrogen number density and $h_{\rm P}$ is Planck's constant.  At $z=5.6$, we find that $\dot{N}_{\rm ion} = 2-4$ photons per hydrogen atom per Gyr in our fiducial model with $\langle \MFP \rangle = 30~h^{-1}\Mpc$.\footnote{Here, the range in $\dot{N}_{\rm ion}$ brackets the thermal histories in Fig. \ref{FIG:sim_thermal_hist}, and we have assumed a spectral index of $\alpha = 2$, consistent with our models from \S \ref{SEC:opacityflucs} and with the fiducial value in \citet{2013MNRAS.436.1023B}.}  This number is revised upwards to $\dot{N}_{\rm ion} = 4-8$ photons per hydrogen atom per Gyr if $\langle \MFP \rangle = 15~h^{-1}\Mpc$.    Such a revision implies that reionization was less photon-starved and therefore of shorter duration than previous studies argued \citep{2007MNRAS.382..325B,2013MNRAS.436.1023B}.

We conclude this section with some important caveats.  Our calculations assume that the quasars shine isotropically and with constant luminosity over the age of the universe.    Since the ionized fraction in the proximity zone responds over the timescale $\GammaHI^{-1}$ to changes in the local radiation intensity, our calculations may over-predict the measurement bias if the luminosities of the quasars in the sample of \citet{2014MNRAS.445.1745W} were, on average, dimmer over a past timespan of $\GammaHI^{-1} \lesssim 50,000$ yrs.  We have also assumed that $\MFP \propto \GammaHI^{2/3}$ throughout the IGM, but this scaling could potentially be shallower in regions of enhanced $\GammaHI$ around quasars.\footnote{This effect can be understood using the simple analytical model of \citet{2000ApJ...530....1M} (see also \citealt{2005MNRAS.363.1031F}; \citealt{2011ApJ...743...82M}; \citealt{munoz14}), in which it is assumed that gas self-shields at densities above the threshold $\Delta_i$.  In this model, 

\begin{equation}
\MFP \propto \int_{\Delta_i}^{\infty}\dd \Delta_b~P(\Delta_b) \propto \Delta_i^{2(\eta-1)/3},
\end{equation}
where $P(\Delta_b)$ is the probability distribution of $\Delta_b$, and the last proportionality uses the power-law approximation $P(\Delta_b)\propto \Delta_b^{-\eta}$ at the relevant densities -- a reasonable approximation for the $P(\Delta_b)$ measured from simulations \citep{2000ApJ...530....1M,2009MNRAS.398L..26B,2011ApJ...743...82M}.     Adopting $\Delta_i \propto \GammaHI^{2/3}$ from the arguments of \citet{2001ApJ...559..507S},  we find that 

\begin{equation}
\MFP \propto \GammaHI^{4 (\eta - 1)/9}.
\end{equation}
Our assumed scaling of $\MFP \propto \GammaHI^{2/3}$ is recovered if $\eta = 2.5$, but this scaling becomes shallower for smaller values of $\eta$.   Thus, if the density PDF is shallower (i.e. $\eta < 2.5$) for the higher-densities that are required for self-shielding at locations close to a quasar, $\MFP$ could scale more weakly with the local $\GammaHI$.}  Compared to our current model, a shallower scaling near quasars would lead to shorter $\MFP$ in those regions, which would reduce the measurement bias.  For these reasons, we caution that the calculations presented above likely represent an upper limit on the effect of quasar proximity zones on direct measurements of the mean free path.  We will quantify these effect in greater detail in future work.    

\begin{figure*}
\begin{center}
\resizebox{18.0cm}{!}{\includegraphics{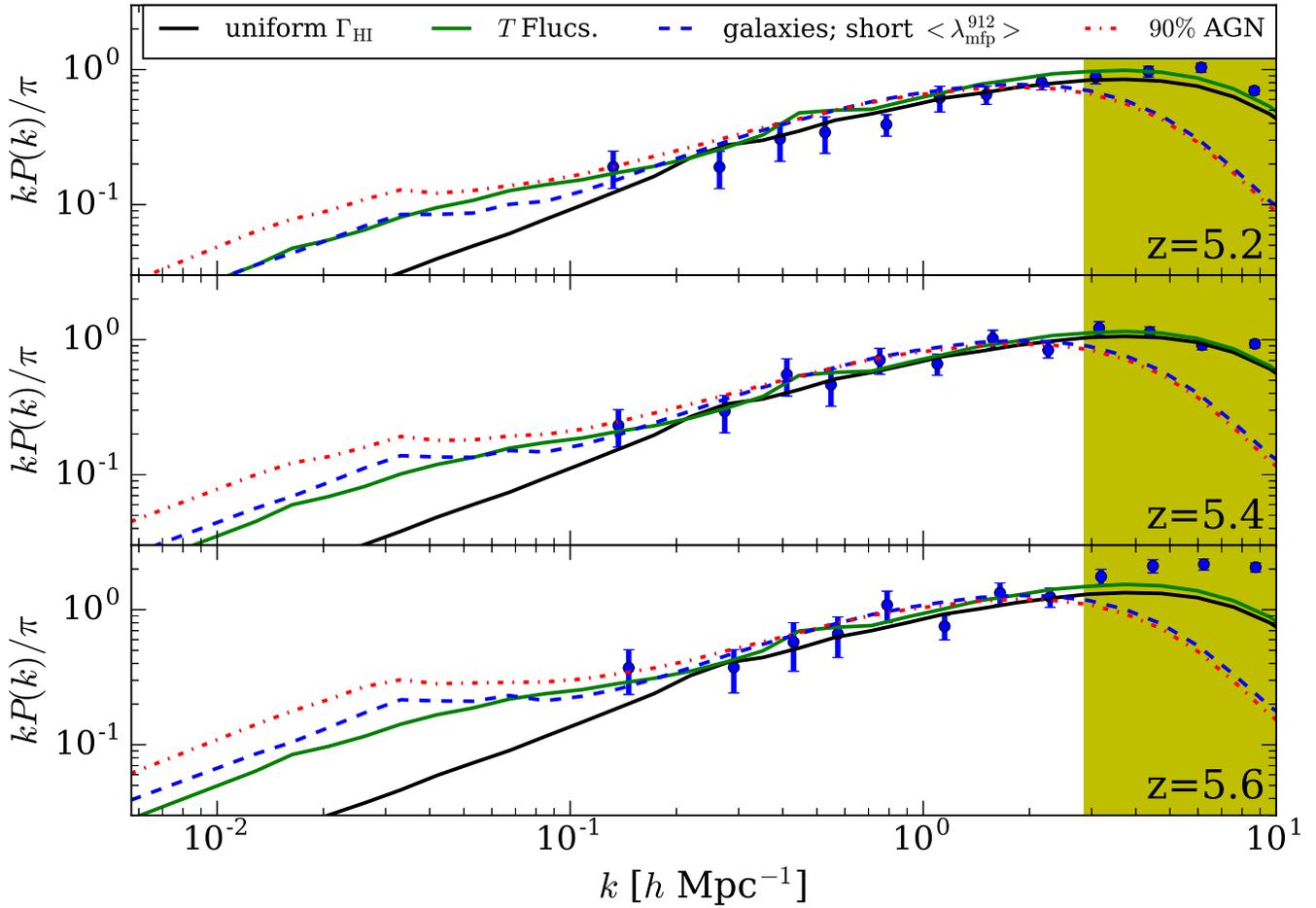}}
\end{center}
\caption{The line-of-sight power spectrum of transmitted flux in the $z>5$ Ly$\alpha$ forest.  The  data points show our measurement of the power spectrum from the $21$ unique quasar spectra in \citet{mcgreer15}.  For reference, the black/solid curve corresponds to a model in which the ionizing background is assumed to be uniform. The blue/dashed curve shows our short mean free path model from \S \ref{SEC:intensityflucs}.  The magenta/dot-dashed curve corresponds to a model from Paper II in which AGN emissions account for $90\%$ of $\langle \GammaHI \rangle$.  The green/solid curves show the model of \citet{2015ApJ...813L..38D} in which the opacity fluctuations are driven by residual temperature fluctuations from the patchy reionization process, instead of by fluctuations in the ionizing background.  All models exhibit enhanced large-scale power over the uniform $\GammaHI$ model at scales of $k \lesssim 0.1~h \Mpc^{-1}$. Future measurements of the flux power spectrum on these scales may rule out density fluctuations alone as the origin of the large-scale opacity fluctuations observed by \citet{2015MNRAS.447.3402B}. }  
\label{FIG:Pk}
\end{figure*}

\section{Comparison to Other Models}
\label{SEC:fluxpower}

\subsection{Current models for the $z>5$ $\taueff$ fluctuations}

{ 
Alternative models have been proposed to account for the large variations in $\taueff$ observed at $z\sim 5.5$. In this section, we will discuss potential ways in which these alternatives might be discriminated from the model considered in this paper.  

So far, three models have been put forth to explain the observations of \citet{2015MNRAS.447.3402B}:

\medskip
\noindent
(i) {\it The short, fluctuating mean free path model of \citet{2015arXiv150907131D}}.  (This is the model considered in the current paper.)

\medskip
\noindent
(ii) {\it The AGN-driven model of \citet{2015arXiv150501853C} (see also \citealt{2016arXiv160608231C} as well as Paper II)}.  In this model, AGN contribute significantly to, or even dominate, the $z\sim 5.5$ ionizing background.  The brightness and rarity of the AGN drive large fluctuations in the ionizing background and in the mean free path\footnote{Note that \citealt{2016arXiv160608231C} and Paper II both assume the same scaling as in the current paper, $\MFP(\boldsymbol{x}) \propto \GammaHI^{2/3}(\boldsymbol{x})/\Delta(\boldsymbol{x})$} .  Unlike the model of \citet{2015arXiv150907131D}, this model does not require a short mean free path to generate large $\taueff$ variations.  In fact, the models presented in \citealt{2016arXiv160608231C} and in Paper II all assume $\langle \MFP \rangle$ values that are consistent with an extrapolation of the mean free path measurements in \citet{2014MNRAS.445.1745W}, e.g. the fiducial model in Fig. \ref{FIG:MFPmodels}.  These authors find that an AGN contribution of $\gtrsim 50 \%$ to the $z\sim 5.5$ \HI\ ionizing background is consistent with the observed $\taueff$ fluctuations.   

\medskip
\noindent
(iii) {\it The reionization temperature fluctuation model of \citet{2015ApJ...813L..38D}}.  In this model the excess opacity fluctuations are caused by residual inhomogeneities in the IGM temperature that were imprinted during the patchy reionization process.  During reionization, intergalactic gas is heated locally to temperatures of $T\sim 20,000-30,000$ K by a passing ionization front.  The gas subsequently cools by the expansion of the universe and by Compton cooling, among other mechanisms.  Large variations in the IGM temperature are generated by different regions being reionized at different times.  These temperature variations result in $\taueff$ fluctuations via the $\sim T^{-0.7}$ dependence of the recombination rate.  \citet{2015ApJ...813L..38D} argued that this model could explain the observed $\taueff$ fluctuations if reionization ends late ($z\approx 6$) and is extended enough, i.e. half-way complete by $z\approx 9$.  This model predicts that the over(under)-dense  regions that are reionized first (last) correspond to the opaque (transmissive) segments of the forest.     
\medskip

Aside from the distribution of Ly$\alpha$ forest $\taueff$ fluctuations, the above models are, in principle, considerably different in their observational signatures.  In model (i), the most opaque segments of the forest correspond to cosmological voids that are under-dense in galactic sources.  In model (iii), the opaque segments correspond to the cold regions that were reionized first, typically the most over-dense regions of the universe.  In model (ii), there is likely less of a correlation between opacity and cosmological density, since the fluctuations are driven by rare AGN sources.  However, the identification of AGN near most large-scale transmissive regions of the forest would be a ``smoking gun" signature of this model.

In the absence of observations that provide definitive evidence for, or against, any of the above models, it is prudent to explore whether additional insights can be gained by applying other statistics to the Ly$\alpha$ forest.}  Following \citet{2015MNRAS.447.3402B}, all studies have used a path length of $L=50~h^{-1}\Mpc$ for $\taueff$.  We first considered varying the pathlength over which $\taueff$ is measured, i.e. measuring $P(<\taueff)$ for $L=10, 25, 75, 100~h^{-1}\Mpc$.  We found that using different $L$ does not bring out differences between our models any more than in the $L=50~h^{-1}\Mpc$ case. Moreover, we note that current data does not provide enough samples to constrain $P(< \taueff)$ for $L> 50~h^{-1}~\Mpc$.  We also considered the distribution of dark gaps and the recently proposed peak height/width distributions of \citet{2016arXiv160503183G}. We found that these statistics are sensitive to the effects of instrumental resolution and noise, which are difficult to remove from current data sets due to their highly inhomogeneous nature, and we additionally found that our models are less reliable for these statistics (see remark at the end of this section).  In this section, we focus exclusively on the line-of-sight power spectrum of the transmitted flux, which is robust to the effects of finite spectral resolution on the scales of interest for our models.  As we will see, the power spectrum offers a complementary view of the high-$z$ Ly$\alpha$ forest; it probes the opacity fluctuations across all scales, but is less sensitive to high $\taueff$ excursions than $P(\taueff)$.         

\subsection{The line-of-sight flux power spectrum}

We present a new measurement of the flux power spectrum at $z=5.2-5.6$ using the sample of 21 quasars presented in \citet{mcgreer15}.  We refer the reader to this paper for details on these spectra. In summary, ten spectra were obtained from the Eschellette Spectrograph and Imager (ESI) on the Keck II telescope, three were obtained with Magellan Eschellette (MagE) spectrograph on the Magellan Clay 6.5m telescope, one was obtained with the Red Channel spectrograph on the MMT 6.5m telescope, and seven were obtained with the X-shooter instrument on the VLT 8m Kueyen telescope.  (For redundant spectra in \citealt{mcgreer15}, we use the spectrum with the higher signal-to-noise.)  Quasar continua were fitted to the broken power law UV continuum model of \citet{2012ApJ...752..162S}.  We conservatively exclude regions of $\Delta v = 10^4~\mathrm{km/s}$ blue-ward of the Ly$\alpha$ emission line to avoid the proximity effect.  We also truncate the spectra at rest-frame $1040$\AA\ to avoid Ly$\beta$ and O{\sc~vi} absorption, as well as uncertainties in the quasar continuum from the presence of Ly$\beta$ and O{\sc~vi} broad emission lines.  This results in somewhat fewer quasars with spectral coverage at $z=5.2$, whereas the $\Delta v = 10^4~\mathrm{km/s}$ cut precludes a reliable measurement of the power spectrum at $z=5.8$ with our spectra.

The one-dimensional flux power spectrum is defined to be 

\begin{equation}
P_F(k) \equiv \frac{1}{L} \left| \tilde{\delta}_F(k) \right|^2.
\end{equation} 
Here, $\delta_F(\lambda) = F(\lambda)/\langle F(\lambda) \rangle - 1$ is the contrast in transmitted flux at wavelength $\lambda$, where $\langle \ldots \rangle$ denotes an average over the total path length $L$,    
and $\tilde{\delta}_F(k)$ denotes its Fourier transform.  We use redshift intervals of $\Delta z = 0.2$ and we assume a piecewise linear interpolation of $\langle F(\lambda) \rangle$ across redshift bins.  We measure logarithmic band powers using the estimator, $P_F(k) = \tilde{\delta}(k)_{\rm raw} \tilde{\delta}(k)_{\rm raw}^* - P_{\rm N}$, where $\tilde{\delta}(k)_{\rm raw}$ corresponds to the raw data, and $P_{\rm N}$ is the noise power spectrum, estimated from the variance in a given redshift bin.   The power spectrum is obtained by an ensemble average over the quasar spectra, in which each spectrum is weighted by $(P_{\rm N} + P_{\rm S})^{-1}$, where ``S" denotes ``signal."  For simplicity we do not account for instrumental resolution, since the finite resolution of our simulations prohibits any meaningful comparison to the measurements on the affected scales.  We also do not correct for contamination by interloping metal lines from lower redshift absorbers, since they are expected to affect $P_F(k)$ at only the $\lesssim 1\%$ level at the redshifts of interest \citep{2013PhRvD..88d3502V}.        

We have tested for the effects of evolution in the mean transmission across a redshift bin and of fluctuations in the quasar continuum over these scales. These effects are the dominant sources of systematic uncertainty for $P_F(k)$ on the largest scales that our measurements probe, $k\approx 0.15~h\Mpc^{-1}$.  To gauge the impact of evolution, we compared our fiducial measurement to one in which $\langle F(\lambda) \rangle$ is assumed to be constant across a redshift bin.  We found an approximately $0.2\sigma$ difference between the two methods for our lowest $k$-bin (and even smaller differences at higher $k$).  To test for continuum fluctuations, we examined a sample of continuum fits for $40$ quasars at $2.1 < z < 4.7$ from \citet{2008A&amp;A...491..465D}.  We found that the continuum power is a factor of $50-100$ lower than $P_F(k)$ in our lowest $k$-bin.  From these tests we conclude that our measurements are robust to the effects of evolution in the mean transmission and continuum fluctuations.   

The data points in Fig. \ref{FIG:Pk} show our measurement of $\Delta_F \equiv k P_F(k)/\pi$.  The error bars represent $1\sigma$ uncertainties obtained by calculating the noise-weighted effective number of modes.  {\addition We compare our measurements to a model that assumes a uniform ionizing background (black/solid), from a higher resolution ($L_{\rm box} = 25 h^{-1}~\Mpc$, $N=2048^3$) simulation}.  We also compare against the short mean free path model from \S \ref{SEC:intensityflucs} (blue/short-dashed), and two alternative models:  (i) {\it The ``AGN-dominated" (90\% AGN) model from Paper II (red/dot-dashed).} In this model, we populate the most massive halos in our simulation with AGN according to a rescaled version of the luminosity function of \citet{2015A&amp;A...578A..83G}.  We tune the escape fraction of galaxies such that the AGN emissions account for $90\%$ of $\langle \GammaHI \rangle$.  This model is otherwise identical to the models in this paper, with $\langle \MFP \rangle(z)$ set to our fiducial values (see Table \ref{TAB:MFPmodels}); (ii) {\it The reionization temperature fluctuation model of \citet{2015ApJ...813L..38D}.}\footnote{We note that the hydro simulations used for the temperature fluctuation model in \citet{2015ApJ...813L..38D} have a significantly higher spatial resolution of $\Delta x = 12.2~h^{-1}\kpc$, compared to the $\Delta x = 97.7~h^{-1}\kpc$ of the larger-scale simulation used for the fluctuating background models in this paper.}    Here we show a model in which reionization begins at $z \approx 13$ and ends at $z\approx6$, with an instantaneous reionization temperature of $T_{\rm reion} = 30,000~\mathrm{K}$ (see the red/solid curves in Fig. 4 of \citealt{2015ApJ...813L..38D}).\footnote{The instantaneous reionization temperature corresponds to the highest temperature achieved behind an ionization front.  Note that $T_{\rm reion} = 30,000~\mathrm{K}$ represents the upper limit of expected temperatures \citep{mcquinn-Xray}.  However, we caution that a similar amplitude of opacity fluctuations may be generated in models with lower $T_{\rm reion}$.   The multi-scale approach of \citet{2015ApJ...813L..38D}, which combines large-scale semi-numerical simulations of reionization with high-resolution hydro simulations, does not capture the correlation between local density and redshift of reionization, $z_{\rm reion}$.  If included, this correlation would increase the amplitude of opacity fluctuations for a given $T_{\rm reion}$. }    The ionizing background is assumed to be uniform in this model. The shaded regions at $k  \gtrsim 3 ~h \Mpc^{-1}$ correspond to scales that are affected by the resolution of our hydro simulation grid, as indicated by the suppressed power in our short mean free path and AGN models. 

All of the models are broadly consistent with the measurements at $k\lesssim 0.3~h \Mpc^{-1}$.  At these scales, the opacity fluctuations are driven by fluctuations in density, as indicated by the agreement with the uniform $\GammaHI$ model.  However, the models begin to diverge from the uniform model on larger scales as the opacity fluctuations transition to being driven by large-scale variations in the ionizing background or temperature.  Although these large-scale variations generate substantially more power at scales corresponding to $k \lesssim 0.2~h\Mpc^{-1}$, we note that $P_F(k)$ in the fluctuating background and temperature models are quite similar.  Thus it will likely be difficult to discriminate between these models using measurements of $P_F(k)$ alone.  Nonetheless, in contrast to $P(< \taueff)$, $P_F(k)$ contains information about the scales on which the opacity variations are no longer driven by density fluctuations.  From the excess power at $k \approx 0.15~h\Mpc^{-1}$ observed in three redshift bins, our measurements disfavor density fluctuations alone as the origin of the opacity fluctuations at the $2\sigma$ level.    Future measurements may improve this limit by extending $P_F(k)$ to lower $k$.  However, this would require larger redshift bins than the $\Delta z = 0.2$ used here, and we caution that quasar continuum uncertainties and redshift evolution could complicate the interpretation of such measurements.    
     
We conclude this section by noting an interesting property of the short mean free path and $90\%$ AGN models, which both have rather low resolutions.   Even though they do not resolve the small-scale features in the forest well (as evident in the premature fall off in power at high $k$), they appear to capture the large-scale properties of the forest fairly robustly.  This is evidenced by our low-resolution simulations matching both our measured $P_F(k)$ and our higher-resolution simulations {\addition (black and green curves)} at intermediate scales, where density fluctuations still dominate ($k=0.2-2~h\mathrm{Mpc}^{-1}$).  This result is similar to our findings in Appendix \ref{SEC:convergence}, and in \citet{2015MNRAS.447.3402B}, that when low-resolution mock spectra are mean flux normalized, they yield $\CPDF$ that are very similar to higher resolution simulations.  While the large-scale predictions of our models are similar, it may be that small-scale statistics can differentiate between them.  Unfortunately, simulations with much higher resolutions are likely necessary to capture small-scale statistics such as the peak height and dark gap statistics.  Thus, there are significant challenges to comparing these models -- which require boxes with $L_{\rm box} \gtrsim 200 h^{-1}\Mpc$, and hence cannot resolve the small-scale forest -- with statistics beyond those considered here.

\section{Conclusion}
\label{SEC:conclusion}

Observations have shown that the dispersion in the Ly$\alpha$ forest $\taueff$ increases quickly at $z>5$, exceeding the dispersion predicted by simple models that assume a uniform ionizing background \citep{2006AJ....132..117F, 2015MNRAS.447.3402B}.  Several models have been proposed to account for these observations \citep{2015ApJ...813L..38D,2015arXiv150907131D, 2016arXiv160503183G, 2016arXiv160608231C}.  In this paper we have explored in further detail the model of \citet{2015arXiv150907131D} for the post-reionization ionizing background.  This model accounts for spatial variations in the mean free path from the enhancement (suppression) of optically thick absorbers where the background is weaker (stronger). 

We showed that accounting for the $z\approx5.5$ opacity fluctuations requires a short spatially averaged mean free path of $\langle \MFP \rangle \lesssim 15 h^{-1}~\Mpc$.  These values are factor of $\gtrsim 3$ shorter than the direct measurement of $\MFP = 44 \pm 7~h^{-1}\Mpc$ at $z\approx 5.2$ \citep{2014MNRAS.445.1745W}.  We further showed that rapid evolution in the mean free path is at odds with our measurement of the global \HI\ photoionization rate, which stays remarkably constant at $4.8 \lesssim z \lesssim 5.6$.   A model in which $\langle \MFP \rangle$ evolves from $\lesssim 15~h^{-1}\Mpc$ at $z=5.6$ to $44~h^{-1}\Mpc$ at $z=5.2$ requires an unnatural factor of $\approx2$ decrease in the ionizing emissivity of the galaxy population in just $\approx100$ million years.   However, we also identified a plausible resolution to this apparent problem.  We showed that direct measurements of the mean free path (from stacking quasar spectra) can be biased higher than $\langle \MFP \rangle$ by up to a factor of $\approx 2$, owing to the enhanced ionizing flux along quasar sight lines (i.e. the ``proximity effect").  It is thus likely that $\langle \MFP \rangle$ is shorter than the measured value $44 h^{-1}~\Mpc$ at $z\approx 5.2$. 

{\addition  We also explored the line-of-sight flux power spectrum as a possible means of testing three models that have been proposed to account for the large opacity variations.  These include the model of \citet{2015arXiv150907131D} explored here, the AGN-dominated scenario of \citet{2016arXiv160608231C} (see also Paper II), and the relic temperature fluctuation model of \citet{2015ApJ...813L..38D}.   Using the 21 quasar sight lines from \citet{mcgreer15}, we found that the large-scale opacity variations are manifest as a $\approx 2\sigma$ excess in power on $k \approx 0.15~h \Mpc^{-1}$ scales, over a model with uniform ionizing background. On these scales, the flux power is quite similar between the models, and all three  are consistent with the data.   The power on scales smaller than those considered here (i.e. $k \gtrsim 3~h\Mpc^{-1}$) may prove to be a more promising avenue for distinguising them.  However, exploring this possibility will require substantial improvements to the current models.  }  

If confirmed, the short $\langle \MFP \rangle$ favored by the model of \citet{2015arXiv150907131D} has implications for reionization and for the structure of the high-$z$ Ly$\alpha$ forest.  In this scenario, reionization is a factor of $\approx 2$ less photon-starved than has been claimed in previous studies, implying a shorter duration of reionization {(all other parameters being fixed)}.   The model also predicts that the most opaque (transmissive) $L\sim 50~h^{-1}\Mpc$ segments of the Ly$\alpha$ forest correspond, on average, to voids (overdensities) where the ionizing background is weaker (stronger).  Interestingly, this is the opposite of what is predicted by the model of \citet{2015ApJ...813L..38D}, which ties the opacity variations to relic temperature fluctuations from reionization.  In that scenario, the overdensities (voids) tend to be the most opaque (transmissive) regions because they are reionized earlier (later); hence they are colder (hotter) by $z\approx 5.5$.  Thus if the model of \citet{2015arXiv150907131D} is correct, it would likely imply that reionization occurred over a short time interval (consistent with the higher ionizing emissivities), minimizing the impact of temperature fluctuations, which would otherwise act to cancel the effect of the background fluctuations.  Conversely, if the model of \citet{2015ApJ...813L..38D} is correct, it would imply a longer mean free path, minimizing the impact of the background fluctuations. A more complete model of the $z>5$ Ly$\alpha$ forest will need to synthesize these effects, as well as possible gas relaxation effects if reionization is ending at $z=5.5-6$.      

At present we are left with three viable models for the large opacity fluctuations, each with very unique implications for \HI\ reionization: 

\medskip

\noindent 
(i) AGN contributed significantly ($\gtrsim 25\%$) to the $z>5$ ionizing background \citep{2015arXiv150501853C, 2016arXiv160608231C}.   This would imply that AGN were more ubiquitous in the $z>5$ Universe than previously thought, and that they contributed more to \HI\ reionization than in most existing theoretical models.  However, in Paper II, we show that these AGN could not have emitted appreciable numbers of \HeII\ ionizing photons or they would have heated up the IGM well above observational bounds \citep{2016arXiv160706467D}. 

\medskip
\noindent
(ii) The mean free path at $z\approx 5.5$ was much shorter than expected, e.g. $\langle \MFP \rangle \lesssim 15h^{-1}$ Mpc if galaxies were the dominant sources of the ionizing background \citep[][and the current paper]{2015arXiv150907131D}.  As noted above, shorter values for the mean free path imply a less photon-starved and quicker reionization process.  In addition, a shorter mean free path could indicate the final stages of the reionization itself.

\medskip
\noindent
(iii) The opacity fluctuations were not driven by the ionizing background, but were instead driven by relic temperature fluctuations from reionization.  This would be the signature of a late-ending ($z\approx6$) and extended reionization process \citep{2015ApJ...813L..38D}.  

\medskip 

All of these models can account for much of the $\taueff$ distribution, but in all models the probability of {\addition occurrance of $\approx 160$ (comoving) Mpc} dark troughs, such as in the sightline to ULAS J0148\verb + 0600 \citep{2015MNRAS.447.3402B}, is low.  Determining which of the above models captures the conditions of the post-reionization IGM will yield new insights into the reionization process.

\section*{Acknowledgements}
{\addition We thank the anonymous referee for helpful comments that improved this manuscript.} The authors acknowledge support from NSF grants AST1312724, AST 1514734 and AST 1614439, as well as NSF XSEDE allocations TG-AST140087 and TG-AST150004.  A.D. thanks Hy Trac for providing his radiative hydrodynamics code. A.D. also thanks Adam Lidz, Brian Siana, George Becker, and Nick Gnedin for useful discussions/correspondence, and Gabor Worseck for providing data on the sample of quasars used in \citet{2014MNRAS.445.1745W}.  The authors thank Ian McGreer, Valentina D'Odorico, and Xiaohui Fan for providing the quasar spectra used for our measurement of the Ly$\alpha$ forest flux power spectrum.   These spectra were collected, in part, at the  European Southern Observatory 
 Very  Large  Telescope, Cerro Paranal, Chile, under programs 084.A-0390, 084.A-0550, 085.A-0299, 086.A-0162, 087.A-0607 and 268.A-5767.
Spectra were also obtained with the MMT Observatory, a joint
facility of the University of Arizona and the Smithsonian Institution,
and with the 6.5-m Magellan Telescopes located at Las Campanas Observatory, Chile.  The authors recognize the cultural and religious importance of the Mauna Kea summit to the indigenous Hawaiian community.   

\bibliography{./master}
\appendix

\section{Numerical Convergence}
\label{SEC:convergence}

In this section we discuss the numerical convergence of our hydro simulation with respect to resolution.  To test this convergence, we ran a suite of smaller simulations at various resolutions, with a fixed box size of $L_{\rm box}=25h^{-1}~\Mpc$.  Like in our production run, we model the reionization of \HI\ in a simplistic way by instantaneously ionizing and heating the boxes uniformly to $20,000$ K at $z=7.5$. We construct $\taueff$ distributions from mock Ly$\alpha$ spectra in the manner described in \S \ref{SEC:methodology}.  {\addition We note that the mock spectra are constructed from $50h^{-1}\Mpc$ sight lines oriented at random angles through our boxes, utilizing periodic boundary conditions. }   At each redshift we rescale the nominal $\GammaHI$ of  the simulation by a constant factor in post-processing to match the observed mean value of $\langle F \rangle_{50}$.  For simplicity, all of the results presented in this section assume a uniform $\GammaHI$.  

\begin{figure}
\begin{center}
\resizebox{8.5cm}{!}{\includegraphics{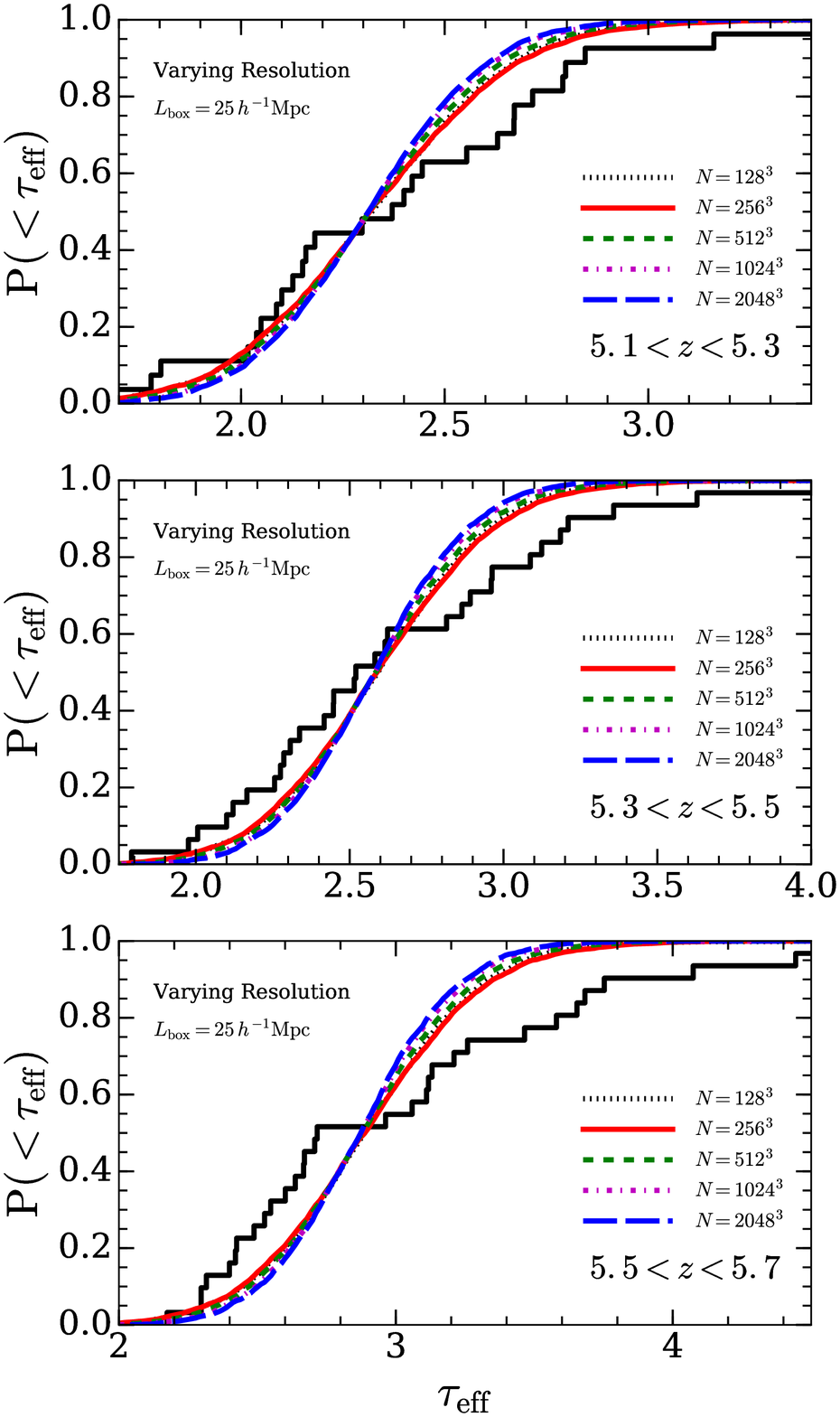}}
\end{center}
\caption{The numerical convergence with respect to simulation resolution for the cumulative probability distribution of $\taueff$.  Here we show results from a suite of simulations with $L_{\rm box} = 25 h^{-1}~\Mpc$ in which the \HI\ photoionization rate is assumed to be uniform.  The black histograms show the observed $\taueff$ distributions of \citet{2015MNRAS.447.3402B}. }
\label{FIG:convergence}
\end{figure}

Figure \ref{FIG:convergence} shows the effect of numerical resolution on $\CPDF$ for three redshift bins.  As before, the black histograms show the observational measurements of \citet{2015MNRAS.447.3402B}.  The curves correspond to different resolutions as indicated in the plot legends.  The $N=256^3$ cases (red/solid) correspond to the same resolution as in our production run (for which $L_{\rm box} =  200h^{-1}~\Mpc$ and $N=2048^3$).  Figure \ref{FIG:convergence} shows that the $\taueff$ distributions at the resolution of our production run are reasonably well converged, with higher resolution runs having a $\CPDF$ that is narrower by a few line widths at most.  

On the other hand, we find that numerically converging on the global photoionization rate is considerably more difficult than for the $\taueff$ distribution.  This difficulty originates from the fact that, for models with uniform $\GammaHI$, the transmission in this highly saturated high-$z$ regime of the forest is dominated by cosmic voids with characteristic under-densities of $\Delta_b \approx 0.2-0.3$.  Table \ref{TAB:simGammaHI} shows for each test simulation the $\GammaHI$ values that have been adjusted to match the observed mean value of $\langle F \rangle_{50}$.  The simulations with lower resolution require a higher $\GammaHI$ because their voids are not as deep as in the simulations with higher resolution ({\addition i.e. the densities of voids in low-resolution simulations are larger}).  Thus, at fixed $\GammaHI$, the lower resolution simulations yield a higher overall opacity than the higher resolution simulations, so $\GammaHI$ must be increased in the former to produce a fixed mean value of $\langle F \rangle_{50}$.  While this has very little impact on our models of the $\taueff$ distribution, we must take it into account in our measurements of the spatially averaged photoionization rate and ionizing emissivity in \S \ref{SEC:globalmeasurements}.  In the next section, we describe how we correct for this finite resolution effect in our measurements.

\ctable[ caption = The Effect of Simulation Resolution on the Mean Transmission of the Ly$\alpha$ Forest.]{c c c c c c c}{

\label{TAB:simGammaHI}

\tnote[*]{All simulations in this table have $L_{\rm box} = 25h^{-1}~\Mpc$.  Values of $\GammaHI$ are given in units of the $\GammaHI$ in the $N_{\rm gas} = 2048^3$ run.   }

}{
\hline\hline
$N_{\rm gas}^{1/3}$ & $\GammaHI^{z=4.8}$ & $\GammaHI^{z=5}$ & $\GammaHI^{z=5.2}$ & $\GammaHI^{z=5.4}$ & $\GammaHI^{z=5.6}$ & $\GammaHI^{z=5.8}$ \\  \hline
\\ [-1ex]
2048 & 1 & 1 & 1& 1 & 1 & 1 \\
1024 & 1.04 & 1.04 & 1.06 & 1.06 & 1.08 & 1.09  \\
512 & 1.17 & 1.18 & 1.21 & 1.22 & 1.26 &1.29   \\
256 & 1.41 & 1.44 & 1.49 & 1.51 & 1.57 & 1.64 \\
128 & 1.47 & 1.47 & 1.5 & 1.52 & 1.58 & 1.53  \\
[1.5ex]
\hline
}

 \section{Measuring the global photoionization rate and ionizing emissivity}
\label{SEC:sim_thermal_hist}

Here we present technical details for our measurements of the spatially averaged photoionization rate, $\langle \GammaHI \rangle$, and ionizing emissivity, $\langle \epsilon^{\rm gal}_{912} \rangle$, presented in \S \ref{SEC:globalmeasurements}.  In the last section we found that the finite resolution of our hydro simulation leads to an overestimate of $\langle \GammaHI \rangle$ (and consequently of $\langle \epsilon_{912} \rangle$).   We have therefore applied redshift-dependent correction factors equal to the ratio $\GammaHI(N=2048^3) / \GammaHI(N=256^3)$, where $\GammaHI(N)$ is the photoionization rate in the test simulation from Appendix \ref{SEC:convergence} with resolution $N$ (see Table \ref{TAB:simGammaHI}).  (Recall that these test simulations have $L_{\rm box} =25h^{-1}~\Mpc$, so the $N=256^3$ run has the same resolution as our $L_{\rm box} = 200 h^{-1}~\Mpc$, $N=2048^3$ production run.)  For example, to correct for resolution effects at $z=5.6$, we divide the $\langle \GammaHI \rangle$ from our production simulation by a factor of $1.57$.    

One of the biggest sources of uncertainty in these measurements is the thermal history of the IGM, as an evolving thermal state could mimick evolution in $\langle \GammaHI \rangle$ and $\langle \epsilon^{\rm gal}_{912} \rangle$.  For a fixed Ly$\alpha$ opacity, hotter IGM temperatures imply lower $\GammaHI$.  We quantify this uncertainty by augmenting the thermal history from our hydro simulation with two semi-analytical models in which the IGM is reionized and heated instantaneously to a temperature of $20,000$ K at $z_{\rm reion}=6$ and $z_{\rm reion} = 10$ (see \citealt{2015arXiv151105992U} for technical details on these models).  These models bracket the range of plausible IGM thermal histories at $5 \leq z \leq 6$, assuming that the heating from \HeII\ reionization is negligible at these redshifts.  This is a reasonable assumption in the standard scenario where galaxies dominate the ionizing background at $z>5$.  

\begin{figure}
\begin{center}
\resizebox{8.8cm}{!}{\includegraphics{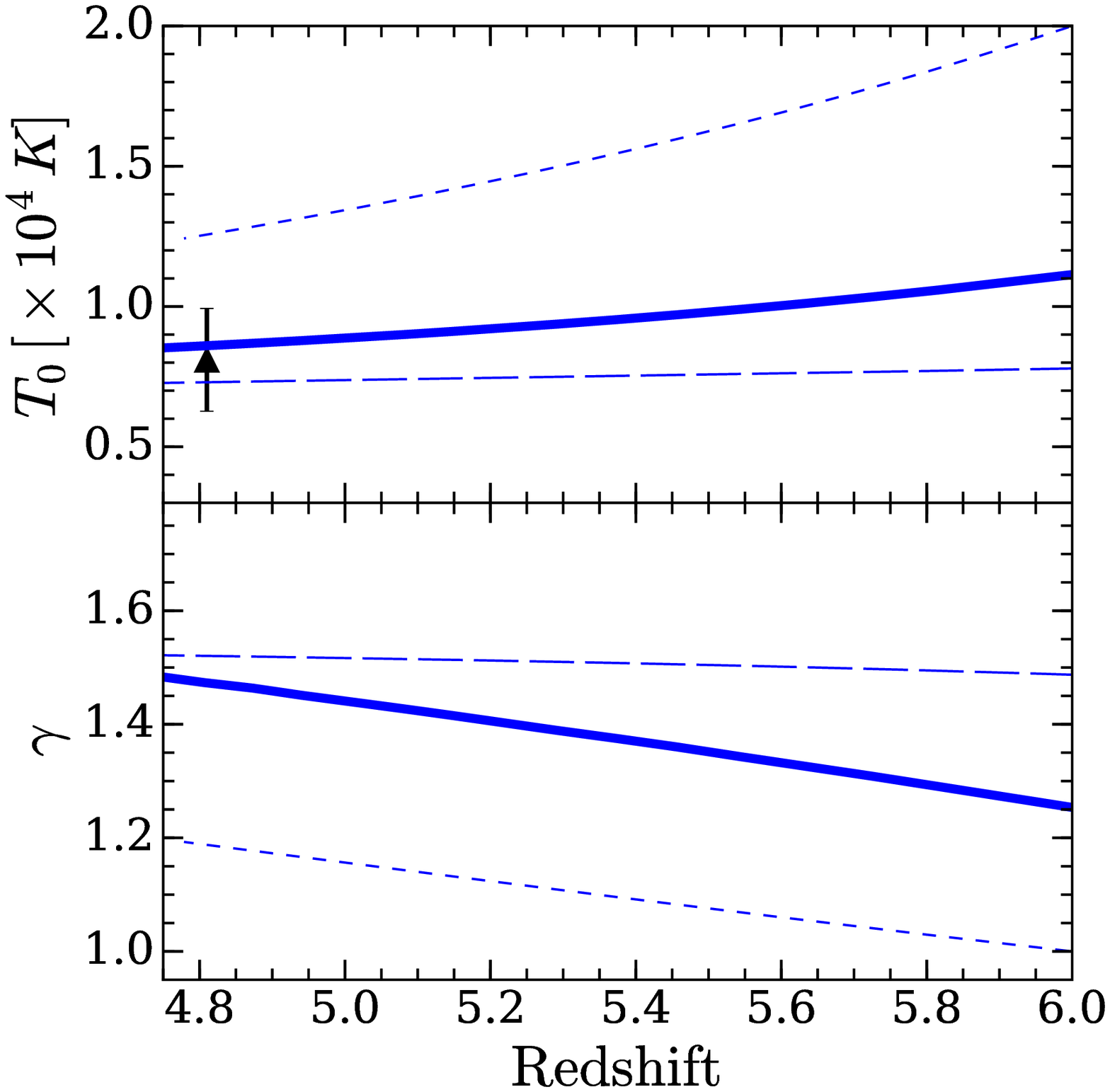}}
\end{center}
\caption{IGM thermal histories used in our measurements of $\langle \GammaHI \rangle$ and $\langle \epsilon_{912}\rangle$ in \S \ref{SEC:globalmeasurements}.  The top and bottom panels show the temperature at the mean density, $T_0$, and the slope parameter of the temperature-density relation, $\gamma$. The thick solid curves show the thermal state in our hydro simulation, in which the IGM is instantaneously reionized and heated to $20,000$ K at $z=7.5$.  For reference, the datapoint at $z=4.8$ shows the highest redshift temperature measurement of \citet{2011MNRAS.410.1096B}, extrapolated to the mean density using the temperature-density relation of our hydro simulation.  The thin curves correspond to thermal histories from a semi-analytical model that we use to quantify the effect of uncertainty in the thermal history.   The thin short-dashed and long-dashed curves show scenarios in which the IGM is instantaneously reionized and heated to $20,000$ K at $z=6$ and $z=10$, respectively.  These models bracket the range of plausible IGM thermal histories after \HI\ reionization, and before the beginning of \HeII\ reionization.       }  
\label{FIG:sim_thermal_hist}
\end{figure}

The thermal state of the IGM is typically parameterized by a temperature-density relation of the form, $T(\Delta) = T_0 \Delta^{\gamma-1}$, where $T_0$ is the temperature at the cosmic mean density, and $\gamma$ specifies the logarithmic slope of the relation \citep{1997MNRAS.292...27H}.   The top and bottom panels of Fig. \ref{FIG:sim_thermal_hist} show $T_0$ and $\gamma$ in our hydro simulation (solid curves) and in our semi-analytic models (short-dashed, $z_{\rm reion} = 6$; long-dashed, $z_{\rm reion} = 10$).  For reference, the datapoint at $z=4.8$ shows the highest redshift temperature measurement of \citet{2011MNRAS.410.1096B}, extrapolated to the mean density using the temperature-density relation of our hydro simulation.  To bracket the effects on $\langle \GammaHI \rangle$ and $\langle \epsilon^{\rm gal}_{912} \rangle$, we rescale our measurements to the thermal histories of our semi-analytic models using the scaling relations of \citet{2013MNRAS.436.1023B}, 

\begin{equation}
\GammaHI(T_0,\gamma) \propto T_0^{-0.575} e^{ 0.7 \gamma}.
\label{EQ:GammaRescale}
\end{equation} 
These limits are depicted by the thin dashed and dotted curves in the top and {\addition middle} panels of Fig. \ref{FIG:galaxy_em}.  We emphasize that these limits bracket only the uncertainty in the thermal history, and do not represent the full range of uncertainties in our measurements.  We note, however, that the former is the dominant source of modeling uncertainty. 

\section{The Effect of Damped Ly$\alpha$ Systems on High-Redshift Opacity Fluctuations }

\label{SEC:DLAs}

Our hydro simulation lacks the resolution and the galaxy formation physics required to model accurately the most optically thick absorbers in the Ly$\alpha$ forest -- the so-called damped Ly$\alpha$ systems (DLAs). In this section we use a simple model to show that DLAs are unlikely to have a significant effect on the distribution of $\taueff$ fluctuations in the $z>5$ Ly$\alpha$ forest.

\begin{figure}
\begin{center}
\resizebox{8.8cm}{!}{\includegraphics{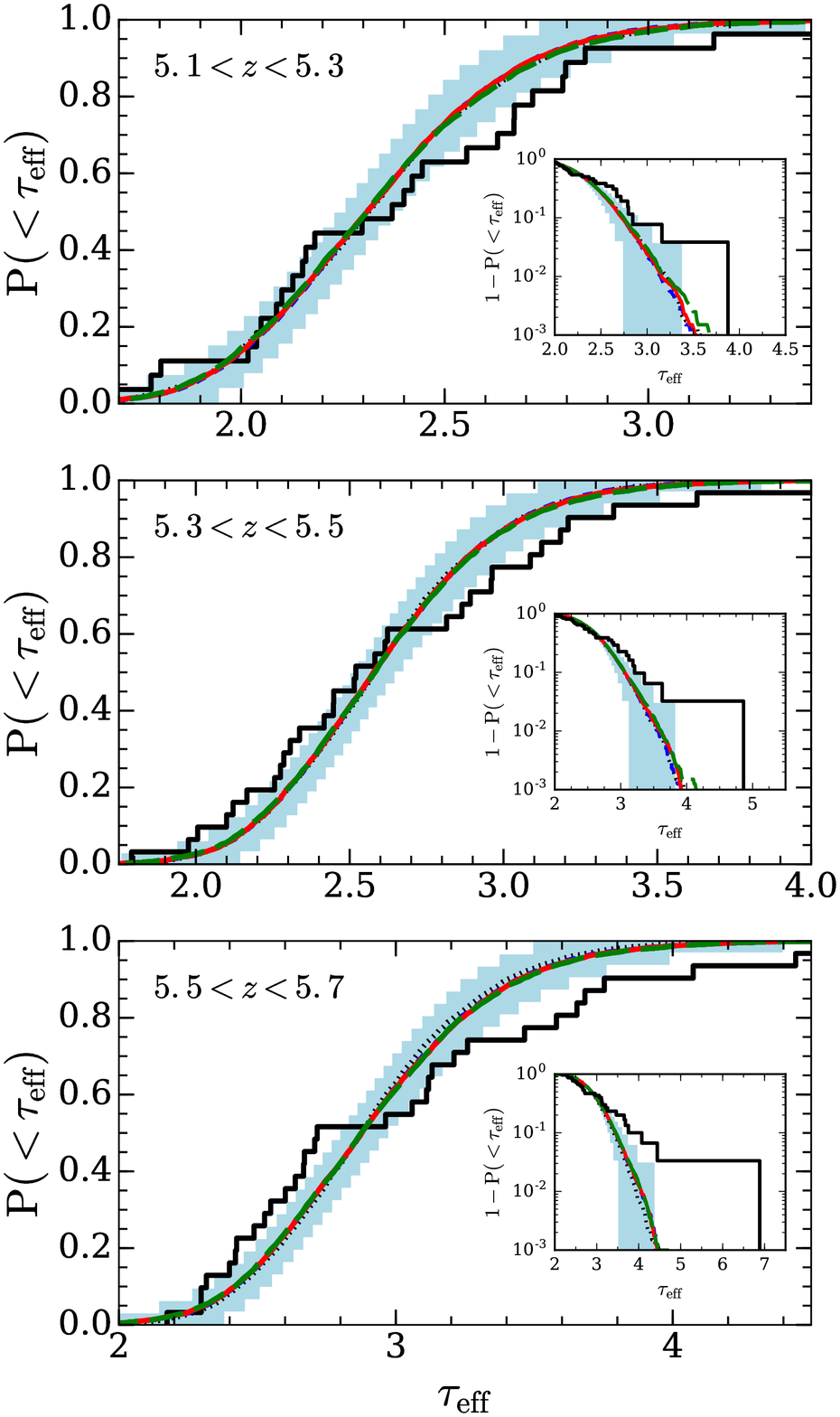}}
\end{center}
\caption{The effect of DLAs on the distribution of Ly$\alpha$ forest opacity fluctuations.  The black histograms show the observational measurements of \citet{2015MNRAS.447.3402B}.  The red/solid curves show a model in which DLAs are drawn from a \HI\ column density distribution that is consistent with observations at lower redshifts (equation \ref{EQ:CDDF}).  The green/long-dashed curves show a model in which the column density distribution is rescaled by a factor of three to boost the abundance of DLAs. The blue/short-dashed and black/dotted curves respectively show our fiducial model from \S \ref{SEC:intensityflucs}, and a uniform $\GammaHI$ model, both of which do not include DLAs.  {\addition All models adopt the fiducial mean free path model in Fig. \ref{FIG:MFPmodels}.}  The blue shading shows 90\% regions estimated by bootstrap sampling of the fiducial model.  These results show that DLAs are unlikely to have a significant effect on the distribution of $\taueff$ fluctuations.  }  
\label{FIG:DLAs}
\end{figure}

We model the population of $z>5$ DLAs by randomly drawing absorbers with column densities $N_{\rm HI} > 10^{19}~\mathrm{cm}^{-2}$ from a column density distribution function (CDDF,  i.e. the number of absorbers per unit column, per unit redshift), $\partial^2 \mathcal{N} / \partial z~\partial N_{\mathrm{HI}} = f(N_{\rm HI}) H_0 (1+z)^2/H(z)$, where $H(z)$ is the Hubble parameter, and 
 
\begin{equation}
\log_{10} f(N_{\rm HI}) = 
\begin{cases}
-1.1 -  \log_{10}( \frac{N_{\rm HI}}{10^{19}~\mathrm{cm}^{-2}})^{1/5}, & \text{if}\ N_{\rm HI} < \\ & 2\times10^{20}~\mathrm{cm}^{-2} \\ \\
-1.1 - \log_{10} 20^{1/5} & \\ - \log_{10}( \frac{N_{\rm HI}}{2\times10^{20} \mathrm{cm}^{-2}}), & \text{otherwise}.
\end{cases}
\label{EQ:CDDF}
\end{equation}
Equation (\ref{EQ:CDDF}) is a simple parametrization that is consistent with observational measurements at a mean redshift of $z=3.7$ \citep[see][and references therein]{2011ApJ...743...82M}.  We note that applying this form to higher redshifts assumes that the CDDF does not evolve significantly with redshift, which is likely a reasonable approximation given the weak evolution in the CDDF observed by \citet{2016MNRAS.456.4488S}.  We generate an ensemble of sight lines, each with $L=50h^{-1}$ Mpc, by randomly throwing down absorbers with Lorentzian line profiles.  We then add the opacities from these absorbers to the sight lines extracted from our simulations (which are presented in the main text). 

Fig. \ref{FIG:DLAs} compares the $\CPDF$ including the absorption from DLAs to those in our fiducial model from the main text (blue/short-dashed), which does not include DLAs. ({\addition Note that all models discussed here adopt the fiducial mean free path model in Fig. \ref{FIG:MFPmodels}})  The red/solid curves correspond to the CDDF of equation (\ref{EQ:CDDF}), while, for the green/long-dashed curves, we rescale $f(N_{\rm HI})$ in equation (\ref{EQ:CDDF}) by a factor of three to explore a scenario with a larger abundance of DLAs. For reference, the black/dotted curves correspond to the uniform $\GammaHI$ model. (Note that all curves in Fig. \ref{FIG:DLAs} are so close that they are indistinguishable.)  In all models, the global $\GammaHI$ has been rescaled to match the observed mean transmission of the forest.  As before, the blue shading shows the 90\% levels estimated by bootstrap sampling of the fiducial model.  These results show that DLAs have a negligible effect on the width of the $\taueff$ distribution.  Even a model in which the DLA abundance is $\approx 3$ times greater than indicated by observations has little impact on the $\taueff$ fluctuations.      

\section{Alternative models for the escape fraction}
\label{SEC:fesc}

The models presented in the main text adopt a constant escape fraction for all galaxies.  In those models, the exponentially declining abundance of halos with mass results in the ionizing background being dominated by ubiquitous faint galaxies near our minimum halo mass limit of $M_{\rm min} = 2\times 10^{10}h^{-1}~\Msun$, corresponding to $M_{\rm AB,1600} \approx -17.5$ at $z\approx 5.5$.  With this magnitude limit, the galaxies are bright enough to be detected in current surveys.  Thus our sources are already much rarer than in models which extrapolate the galaxy luminosity function below detection limits.  Nonetheless, in this section we will test whether the $z\approx 5.5$ $\taueff$ dispersion can be reproduced by varying our assumptions about $f_{\rm esc}$.  In particular, we will consider toy models for the escape fraction that increase the contribution from yet rarer and brighter galaxies to the ionizing background.  

Recall that the Lyman break in the spectra of galaxies is parameterized in our simulations by $f_{\rm esc}/A_{912}$, where $A_{912} = L_{\nu}(1600\mathrm{\AA})/L_{\nu}(912 \mathrm{\AA})$ represents the intrinsic Lyman break due to absorption in stellar atmospheres, and $f_{\rm esc}$ represents the escape fraction of \HI\ ionizing photons into the IGM.  The parameters $f_{\rm esc}$ and $A_{912}$ are completely degenerate; only their ratio, i.e. the overall break, is fixed by matching the mean $\langle F \rangle_{50}$ to observations.  Thus extracting a value for $f_{\rm esc}$ requires specifying $A_{912}$, which depends on the star formation history of the galaxy.  In addition, the value of $f_{\rm esc}$ depends on the highly uncertain thermal state of the IGM.  Hotter temperatures will yield lower $f_{\rm esc}$ and vice versa.  Here we shall adopt $A_{912} = 2$, which is approximately the lower limit of expected values for young stellar populations \citep{1999ApJS..123....3L}.  We will also rescale our results to the hotter thermal history in Fig. \ref{FIG:sim_thermal_hist}.  With these choices, the values of $f_{\rm esc}$ quoted in this section can be thought of as rough lower limits for the escape fraction.  

\begin{figure}
\begin{center}
\resizebox{8.5cm}{!}{\includegraphics{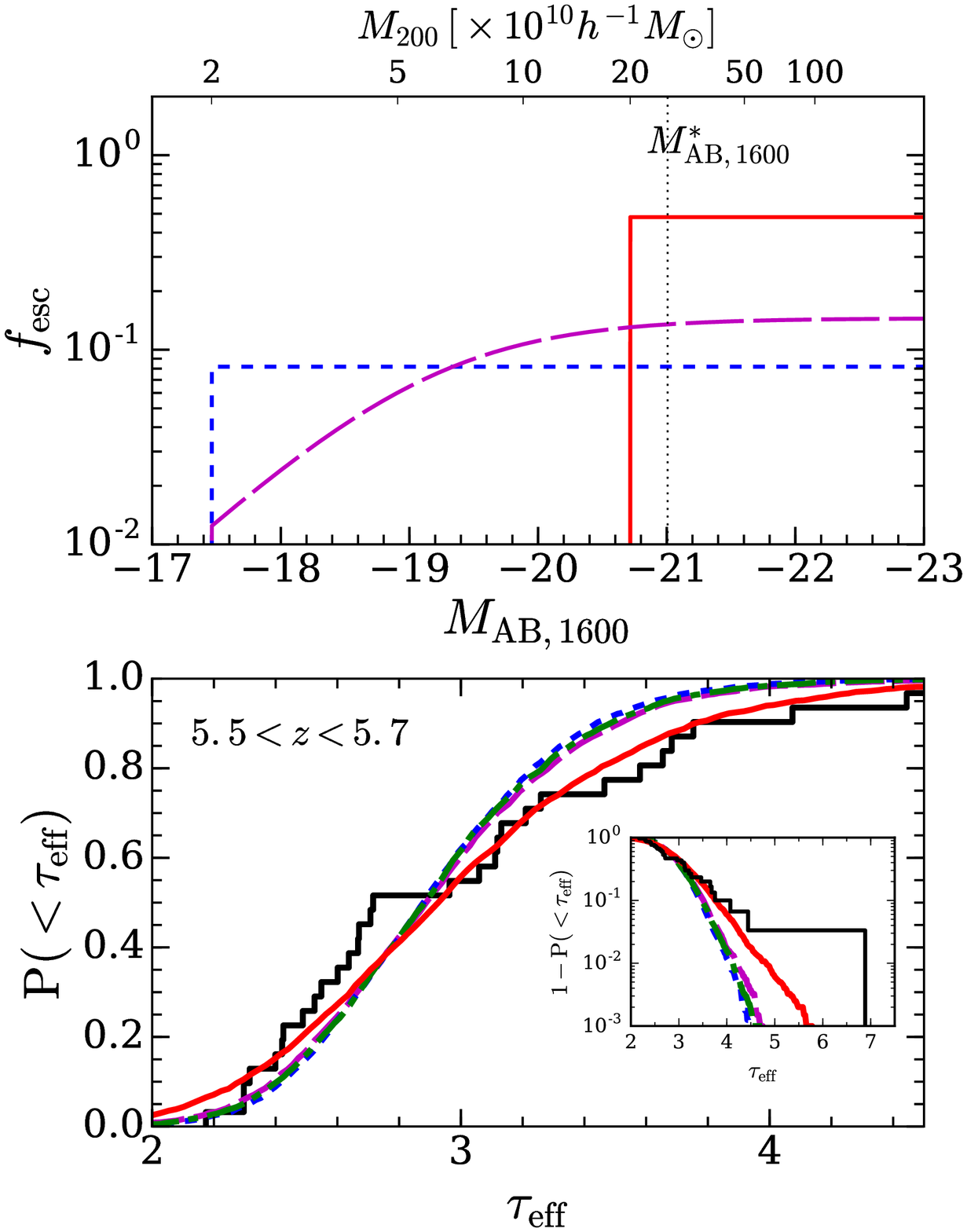}}
\end{center}
\caption{Ly$\alpha$ forest opacity fluctuations in alternative models for $f_{\rm esc}$.  The curves in the top panel correspond to lower limits on $f_{\rm esc}$ in our models (see text in Appendix \ref{SEC:fesc}). In all models we assume $\langle \MFP  \rangle = 30~h^{-1}\Mpc$.  The vertical dotted line in the top panel shows $M^*_{\rm AB, 1600}$ for the luminosity function measurement of \citet{2015ApJ...803...34B}.  The blue/short-dashed curves show our fiducial model from the main text.  The green/long-dashed curve (barely distinguishable from the blue/short-dashed curve) in the bottom panel corresponds to a model in which a random $1\%$ of galaxies are given $f_{\rm esc} = 1$.  {\addition The magenta/long-dashed and red/solid curves correspond to models in which the ionizing background is dominated by rarer, more biased galaxies.}   }
\label{FIG:fesc}
\end{figure}

The top panel of Fig. \ref{FIG:fesc} shows $f_{\rm esc}$ as a function of absolute magnitude for three out of the four toy models considered in this section.  The top axis shows the corresponding halo masses obtained by abundance matching (see main text).  The blue/short-dashed and red/solid curves correspond to constant $f_{\rm esc}$ models. In the former we set $M_{\rm min} = 2\times10^{10}h^{-1}~\Msun$, the same as in our fiducial models, whereas in the latter we set $M_{\rm min} = 2\times10^{11}h^{-1}~\Msun$.  The magenta/long-dashed curve corresponds to an intermediate model where $f_{\rm esc}$ increases with luminosity.  Below, we also consider a model in which we allow a random fraction of the galaxy population to have $f_{\rm esc} =1$.  This models are not represented in the top panel of Fig. \ref{FIG:fesc}.  

The bottom panel of Fig. \ref{FIG:fesc} shows $\CPDF$ at $z=5.6$ for these models, assuming our fiducial value of $\langle \MFP \rangle = 30 h^{-1}~\Mpc$.  Let us first consider the green/dot-dashed curve (hardly distinguishable from the blue/short-dashed one), representing a model in which a random 1\% of galaxies, corresponding to $2,108$ halos in our simulation box, are given $f_{\rm esc} = 1$.  The $\taueff$ fluctuations in this model are not significantly larger than those in the fiducial model (blue/short-dashed).  This is because the background is still dominated by the faint galaxies near the $M_{\rm min}$ threshold, and the small fraction of galaxies with $f_{\rm esc} = 1$ are typically not bright enough to introduce substantial large-scale fluctuations in the background.  The magenta/long-dashed curve is also quite similar to the blue/short-dashed curve. Despite the fact that $f_{\rm esc}$ increases steeply with luminosity in this model, the ionizing background still receives a large contribution from the substantial population of $51,854$ halos with $M_{200} \gtrsim 5\times10^{10}h^{-1} ~\Mpc$ in our simulation box.   

The amplitude of $\taueff$ fluctuations in the red/solid model is a significantly better match to the observations.  In this model, the ionizing background is sourced entirely by the mere $4,040$ halos in our box with $M_{200} \gtrsim 2\times10^{11}h^{-1}~\Msun$.  However, the top panel shows that achieving the required background strength with this small source population requires that $f_{\rm esc} \gtrsim 0.5$.  Such large mean escape fractions for the most luminous galaxies are difficult to reconcile with recent observations \citep[e.g.][]{2010ApJ...725.1011V, 2015ApJ...810..107M, 2015ApJ...804...17S, 2016A&amp;A...585A..48G}.  We emphasize also that the values of $f_{\rm esc}$ quoted here are conservatively low, owing to our choices for $A_{912}$ and for the thermal state of the IGM. We thus conclude that matching the observed $\taueff$ fluctuation amplitude with these models requires that the ionizing background is dominated by extremely rare galaxies with implausibly large $f_{\rm esc}$.

\end{document}